\documentclass{elsart}
\usepackage{graphicx,harvard,amssymb}

\makeatletter
\def\elsartstyle{%
	\def\normalsize{\@setfontsize\normalsize\@xiipt{14.5}}
	\def\small{\@setfontsize\small\@xipt{13.6}}
	\let\footnotesize=\small
	\def\large{\@setfontsize\large\@xivpt{18}}
	\def\Large{\@setfontsize\Large\@xviipt{22}}
	\skip\@mpfootins = 18\p@ \@plus 2\p@
	\normalsize
}
\makeatother

\def\url#1{{\ttfamily\def\/{/\discretionary{}{}{}}#1}}

\begin{document}

\newcommand{\ttbs}{\char'134}
\newcommand{\AmS}{{\protect\the\textfont2
  A\kern-.1667em\lower.5ex\hbox{M}\kern-.125emS}}

%put here your newcommands
\newcommand{\beq}{\begin{equation}}
\newcommand{\qmatrix}[1]{\left[\matrix{#1}\right]}
\newcommand{\eeq}{\end{equation}}
\newcommand{\un}{\underline}
\newcommand{\unun}[1]{\underline{\underline{#1}}}
\newcommand{\beqa}{\begin{eqnarray}}
\newcommand{\eeqa}{\end{eqnarray}}
\newcommand{\Npix}{N_{\rm pix}}
\newcommand{\Spix}{\sigma_{\rm pix}}
\newcommand{\aj}{AJ}
\newcommand{\prd}{Phys. Rev. D}
\newcommand{\araa}{ARA\&A}
\newcommand{\apj}{ApJ}
\newcommand{\apjs}{ApJS}
\newcommand{\aap}{A\&A}
\newcommand{\aaps}{A\&AS}
\def\farcm{\hbox{$.\mkern-4mu^\prime$}}
\def\etal{{et al.\/}}
\def\la{\mathrel{\mathchoice {\vcenter{\offinterlineskip\halign{\hfil
$\displaystyle##$\hfil\cr<\cr\sim\cr}}}
{\vcenter{\offinterlineskip\halign{\hfil$\textstyle##$\hfil\cr
<\cr\sim\cr}}}
{\vcenter{\offinterlineskip\halign{\hfil$\scriptstyle##$\hfil\cr
<\cr\sim\cr}}}
{\vcenter{\offinterlineskip\halign{\hfil$\scriptscriptstyle##$\hfil\cr
<\cr\sim\cr}}}}}
\def\ga{\mathrel{\mathchoice {\vcenter{\offinterlineskip\halign{\hfil
$\displaystyle##$\hfil\cr>\cr\sim\cr}}}
{\vcenter{\offinterlineskip\halign{\hfil$\textstyle##$\hfil\cr
>\cr\sim\cr}}}
{\vcenter{\offinterlineskip\halign{\hfil$\scriptstyle##$\hfil\cr
>\cr\sim\cr}}}
{\vcenter{\offinterlineskip\halign{\hfil$\scriptscriptstyle##$\hfil\cr
>\cr\sim\cr}}}}}
\begin{frontmatter}
\title{The Sky Polarization Observatory (SPOrt)}
\author[BO]{S.~Cortiglioni},
\author[BO]{G.~Bernardi},
\author[BO]{E.~Carretti},
\author[BO]{L.~Casarini},
\author[BO]{S.~Cecchini},
\author[BO]{C.~Macculi},
\author[BO]{M.~Ramponi},
\author[BO]{C.~Sbarra},
\author[IRA]{J.~Monari},
\author[IRA]{A.~Orfei},
\author[IRA]{M.~Poloni},
\author[IRA]{ S.~Poppi},
\author[MI1]{G.~Boella},
\author[MI1]{S.~Bonometto},
\author[MI1]{L.~Colombo},
\author[MI1]{M.~Gervasi},
\author[MI1]{G.~Sironi},
\author[MI2] {M.~Zannoni},
\author[TO]{M.~Baralis},
\author[TO]{O.A.~Peverini},
\author[TO]{R.~Tascone},
\author[TO]{G.~Virone},
\author[FI1]{R.~Fabbri},
\author[FI2]{V.~Natale},
\author[PA]{L.~Nicastro},
\author[NG]{K.--W.~Ng},
\author[RU1]{E.N.~Vinyajkin},
\author[RU1]{V.A.~Razin},
\author[RU2]{M.V.~Sazhin },
\author[RU3]{I.A.~Strukov},
\author[ASI]{B.~Negri}

\address[BO]{IASF--CNR Bologna, via P. Gobetti 101, I--40129 Bologna, Italy}
\address[IRA]{IRA--CNR Bologna, Via Bassini 15, I--20133 Milano, Italy}
\address[MI1]{Phys. Dep. G. Occhialini, Universit\`a di Milano­-Bicocca, 
     Piazza della Scienza 3, I20126 Milano, Italy }
\address[MI2]{ IASF--CNR Milano, Via Bassini 15, I--20133 Milano, Italy}
\address[TO]{IEIIT--CNR Torino, c.so Duca degli Abruzzi 24, I--10129 Torino, 
   Italy}
\address[FI1]{Dipartimento di Fisica, Universit\`a di Firenze, 
   via Sansone 1, I--50019 Sesto Fiorentino, Italy}
\address[FI2]{IRA--CNR Firenze, Largo E. Fermi 5, I--50125 Firenze, Italy}
\address[PA]{IASF--CNR Palermo, Via U. La Malfa 153, I--90146 Palermo, Italy}
\address[NG]{Institute of Physics \& ASIAA, Academia Sinica, Taipei, 
   Taiwan 11529}
\address[RU1]{NIRFI, 25 B.Pecherskaya st., Nizhnij Novgorod 603600/GSP--51, Russia}
\address[RU2]{Schternberg Astronomical Institute,Moscow State University, 
      Moscow 119992, Russia}
\address[RU3]{Space Research Institute (IKI), Profsojuznaja ul. 84/32, Moscow 
     117810, Russia}
\address[ASI]{Agenzia Spaziale Italiana, Viale Liegi 26, I--00198 Roma, Italy}
\begin{abstract}
SPOrt is an ASI--funded  experiment specifically 
designed to measure the sky polarization at 22, 32 and 90 GHz, which 
was selected in 1997 by ESA to be flown on the International Space Station. 
Starting in 2006 and for at least 18 months, it will be taking direct 
and simultaneous measurements of
the Stokes parameters Q and U at 660 sky pixels, with ${\rm FWHM}=7^{\circ}$.
Due to development efforts over the past few years,
the design specifications have been significantly improved
with respect to the first proposal. 
Here we present an up--to--date description of the instrument, which 
now warrants a pixel sensitivity of 1.7 $\mu$K for the polarization of 
the cosmic background radiation, assuming two years of observations.
We discuss SPOrt scientific goals in the light of WMAP results, in particular 
in connection with the 
emerging double--reionization cosmological scenario.
\end{abstract}


\end{frontmatter}

%
%________________________________________________________________

\section{Introduction}

The last decade of the past century has seen a number of great
achievements in observational cosmology. Among them, the assessment of
the nature of the cosmic microwave background radiation (CMBR), by
COBE (Mather \etal\ 1990, 1999), and the detection of its tiny temperature 
fluctuations (Smoot \etal\ 1992; Bennett \etal\ 1996), providing
a sound confirmation of the gravitational instability picture.
After COBE, a number of ground and balloon experiments 
(see, e.g., De Bernardis \etal\ 2000, Hanany \etal\ 2000 for a review),
together with the ongoing WMAP mission (Bennett \etal\ 2003),
 extended the measured angular
spectrum of temperature fluctuations up to $l \simeq 800$.
By fitting model predictions to these measurements, the
multi--dimensional parameter space of available cosmological models
was significantly constrained.

The main target of further analysis was clearly polarization. Besides
conveying independent information on the cosmological model, so as to
remove residual degeneracies between various parameters, large angle
polarization data allow to explore the thermal history of the Universe
at low redshift, in the epoch when primeval fluctuations gradually
turned into today's systems and objects. Upper limits to the
polarization signal had been put by ground experiments
 (see, e.g., Keating \etal\ 2001; Hedman \etal\ 2002), while DASI
(Kovac \etal\ 2002) claimed a detection with 
significance of $4.9 \sigma$ on angular
scales $\sim 0.5^o$.

The entry into the new century of CMBR exploration was however set by
WMAP detecting a clear signal of TE correlation at large angular 
scales (Kogut \etal\ 2003).
This was much more than an aid to remove parameter degeneracies.
Its intensity carried unexpectedly important information
on the history of our Universe, operating a radical selection among
reionization scenarios.

This finding makes the SPOrt experiment quite timely. This experiment,
fully funded by the Italian Space Agency (ASI), was proposed to the European Space 
Agency (ESA) in 1997 to put a polarimeter onboard the International Space
Station (ISS). Its frequency range (22--90~GHz) and angular resolution
(${\rm FWHM}=7^{\circ}$), together with an almost full--sky coverage, are optimal 
to search for CMBR polarization on large angular scales. 
This strategy has now received an unexpectedly
strong support by the first--year WMAP release, which favours
values of the Universe optical depth $\tau \sim 0.17$.

Such high values need a confirmation, which shall come first by further
WMAP releases. An independent confirmation, based on an instrument
devoted to polarization and {\it operating independently from anisotropy},
will then be allowed by the SPOrt experiment. It should be stressed that
a safe assessment on a so large $\tau$ is not only quantitatively
relevant, but changes our views on the history of cosmic reionization
and on the very epoch when the first objects populated the Universe.
Besides allowing to measure $\tau$,
large angle polarization will also put new constraints on the nature
of Dark Energy (DE), which determines the relation between time and
scale factor in the epoch when reionization occurred.

Only full sky experiments, based on space, can provide information
on large angular scale polarization, ground based or balloon experiments
being intrinsically unable to access such scales.

In the light of such considerations, here we report on the present
advancement of the SPOrt experiment, on the technical aspects of the
polarimeter we shall fly and on the physics we expect to explore
through its observations.

The plan of the paper is as follows: Sect.~\ref{thry} reviews 
large--scale CMBR polarization theory as well as the current 
scenario, including foregrounds; 
Sect.~\ref{sport} describes the SPOrt experiment, with special emphasys 
on its technological and design aspects;
Sect.~\ref{phys}  is focused on the main physics results SPOrt is expected to
provide, compared to WMAP. The conclusion 
are summarised in Sect.~\ref{conc}.

\section{Scientific targets: theory and current scenario}
\label{thry}
\subsection{Anisotropy and polarization: definitions}

SPOrt aims at measuring the polarization of CMBR. Quite in general, to
define the properties of linearly polarized radiation coming from a
sky direction $\hat n$, we make use of a basis $(\hat e_1,\hat e_2)$,
in the plane perpendicular to $\hat n$, to define the $2\times 2$
tensor $I_{ij} = \langle E_i E_j \rangle / 4\pi c$ ($E_i$ are
components of the electric field vector $\hat E$, perpendicular
to $\hat n$). The total intensity of the radiation from the direction 
$\hat n$ is then the tensor trace $I = I_{11} + I_{22}$. By angle averaging,
we obtain the {\it average} radiation intensity $\bar I$.

This average value ought to be subtracted from the signal, to
detect the main CMBR features. We therefore replace $I_{ij}$ by the
tensor $\Delta_{ij} = {I_{ij}/{\bar I}} - \delta _{ij}/2$.
Its components are directly related to the temperature anisotropies
and the Stokes parameters accounting for linear polarization, as follows:
\begin{equation}
\label{deltas}
T=(\Delta _{11}+\Delta _{22})/4
~,~Q=(\Delta _{11}-\Delta _{22})/4
~,~U=\Delta _{12}/2.
\end{equation}

$Q$ and $U$ depend on
the basis $(\hat e_1,\hat e_2)$. If such basis is rotated
by an angle $\psi $, so that
${\hat e_1}^{\prime }=\cos \psi \ {\hat e_1}+\sin \psi \ {\hat e_2}$
and
${\hat e_2}^{\prime }=-\sin \psi \ {\hat e_1}+\cos \psi \ {\hat e_2}$,
the Stokes parameters change into
$Q^{\prime }= \cos 2\psi \ Q+\sin 2\psi \ U$,
$U^{\prime }=-\sin 2\psi \ Q+\cos 2\psi \ U$.
According to Tegmark \& de Oliveira--Costa (2001), it is then convenient to 
define a 3--component quantity
\begin{equation}
\label{dtqu}
\mathbf{D} \equiv (T,Q,U)
\end{equation}
which transforms into
\begin{equation}
\label{rotation}
\mathbf{D^{\prime }}=\mathbf{R}(\psi )\mathbf{D}~,
\end{equation}
when the basis is rotated by $\psi$; here
\begin{equation}
\label{rpsi}
\mathbf{R}(\psi ) =
\left( \matrix {1 &     0      &     0      \cr
                0 & \cos 2\psi & \sin 2\psi \cr
                0 &-\sin 2\psi & \cos 2\psi \cr}
\right)
\end{equation}
is a $3 \otimes 3$ matrix.

\subsubsection{Angular Power Spectra}
\label{defaps}

$T$, $Q$ and $U$ depend on the sky direction $\hat n$. From them
we obtain angular power spectra (APS), by expanding on
suitable spherical harmonics. The spherical harmonics currently
used to expand anisotropies ($T$) are however inappropriate for
polarization, as the combinations $Q \pm iU$ are quantities
of spin $\pm 2$ (Golberg \etal\ 1967). In this case, one makes
use of the spin--weighted harmonics $\,_{\pm 2}Y_l^m$ 
(Zaldarriaga \& Seljak 1997) %%%SB
\begin{equation}
\begin{array}{rll}
\label{expansions}
T(\hat n)& = &\sum_{lm} a_{T,lm} Y_{lm}(\hat n) \\
(Q+iU)(\hat n) & = & \sum_{lm} a_{+2,lm}\; _{+2}Y_{lm}(\hat n) \\
(Q-iU)(\hat n) & = & \sum_{lm} a_{-2,lm}\; _{-2}Y_{lm}(\hat n). \\
\end{array}
\end{equation}
In order to relate CMBR polarization to its origin, instead of
using the Stokes parameters $Q$ and $U$,
it is more convenient to describe linear polarization through
two suitable {\it scalar} quantities $E$ and $B$ (see: Seljak \& 
Zaldarriaga 1997 and Kamionkowski \etal\ 1997a,
who give a slightly different notation). %%%SB
Theory provides direct predictions on such quantities
and the treatment in terms of $E$ and $B$ takes full account of
the spin--$2$ nature of polarization. 
Their scalar harmonics coefficients 
are related to $a_{\pm 2,lm}$ by the linear combinations
$a_{E,lm} = -(a_{2,lm}+a_{-2,lm})/2 $
and          
$a_{B,lm} = (a_{2,lm}-a_{-2,lm})/2i$.

These
combinations transform
differently under parity. Thanks to that, just four 
angular spectra $C_{Yl}$
($Y=T$, $E$, $B$ and $TE$) characterize fluctuations, {\it if
they are Gaussian}. These APS are the transforms 
of the autocorrelations of $T$,
$E$ and $B$ and the cross--correlation between $E$ and $T$; invariance
under parity grants the vanishing of the cross--correlations
between $B$ and any other quantity, and the related APS. More
in detail, such APS are defined as follows:
\begin{equation}
\label{apsdefine}
\begin{array}{lllclll}
\langle a_{T,lm}^{*} a_{T,lm^{\prime }}\rangle & = &
\delta_{m,m^{\prime}} C_{Tl}; &\,\,\,\,\,\,\,\, &
\langle a_{E,lm}^{*} a_{E,lm^{\prime }}\rangle & = &
\delta_{m,m^{\prime}} C_{El}\\
\langle a_{B,lm}^{*} a_{B,lm^{\prime }}\rangle & = &
\delta _{m,m^{\prime}}C_{Bl}; &\,\,\,\,\,\,\,\, &
\langle a_{T,lm}^{*} a_{E,lm^{\prime }}\rangle & = &
\delta_{m,m^{\prime}} C_{TEl}.
\end{array}
\end{equation}

In order to fit data with models, one must also be able 
to invert these relations, working out signal correlations from
APS. This operation can be fairly intricate, because of the
complex properties of spherical harmonics. One can start the
procedure by defining the following functions:
\begin{eqnarray}
\label{f2lm}
F_{1,lm}(\theta) & = & 2 \sqrt{\frac{(l-2)!(l-m)!}{(l+2)!(l+m)!}}
\left\{ \left[\frac{m^2-l}{\sin^2\theta} -
\frac{l(l-1)}{2}\right]\cdot \right. \nonumber \\
& & \left. P_l^m(\cos \theta) + 
(l+m)\frac{\cos\theta}{\sin^2\theta}P_{l-1}^m(\cos\theta) \right\} \nonumber \\
F_{2,lm}(\theta) & = & 2 \sqrt{\frac{(l-2)!(l-m)!}{(l+2)!(l+m)!}}\cdot
\frac{m}{\sin^2\theta} \cdot \big[ -(l-1)\cdot  \nonumber \\
& &  \cos\theta P_l^m(\cos\theta) + 
(l+m) P_{l-1}^m(\cos\theta) \big].
\end{eqnarray}
Then, simple but lengthy calculations provide relations,
similar to the equation
\begin{equation}
\label{t1t2}
\langle T(1)~T(2)\rangle =
\sum_l{\frac{2l+1}{4\pi }}C_{Tl}P_l(\cos \alpha_{12})
\end{equation}
($\alpha_{12}$ is the angle between the directions 1 and 2),
relating fluctuations to spectra, for Stokes parameters and their
correlations. Let us remind that the l.h.s. of Eq.~(\ref{t1t2})
prescribes an averaging operation which, in principle, is an
ensemble average. To work out such function from data, the
average is taken on all pairs of sky directions with angular 
separation $\alpha_{12}$. 

The relations analogous to Eq.~(\ref{t1t2}), for the Stokes
parameters, are simpler if the basis vector $\hat e_1$
is aligned with the line connecting the points 1 and 2. 
There they read
\begin{eqnarray}
\label{t1u2}
\langle Q(1)Q(2) \rangle & = & \sum_l {2l+1 \over 4 \pi} [C_{El}
F_{1,l2}(\alpha_{12})-C_{Bl} F_{2,l2}(\alpha_{12})] \nonumber \\  
\langle U(1)U(2) \rangle & = & \sum_l {2l+1 \over 4 \pi} [C_{Bl}
F_{1,l2}(\alpha_{12})-C_{El} F_{2,l2}(\alpha_{12}) ]\nonumber \\
\langle T(1)Q(2) \rangle & = & -\sum_l {2l+1 \over 4 \pi}C_{TEl}
F_{1,l0}(\alpha_{12}) \nonumber \\
\langle T(1)U(2) \rangle &=& 0.
\end{eqnarray}
For a generic basis, such relations are more complicated
(Ng \& Liu 1999).
Owing to their definition and to Eq.~(\ref{rotation}), however, it is 
evident that the correlation are components of the $3 \otimes 3$ matrix
$\langle D_i D_j \rangle $, which transforms as follows:
\begin{equation}
\label{didi}
\langle D_{i_1}^{\prime } D_{i_2}^{\prime } \rangle =
R_{i_1,j_1}(\psi_1) \langle D_{j_1}D_{j_2} \rangle R_{i_2,j_2}(\psi_2)
\end{equation}
with $\mathbf{R}$$(\psi )$ defined in Eq.~(\ref{rpsi}); $\psi_{1,2}$
are the rotation angles to put the local polar basis in 1,2 along the
great circle connecting the pair.
Making use of these transformations, we can
pass from the priviledged basis, where Eqs.~(\ref{t1u2}) hold,
to the generic basis needed for observational or modellization aims.

When real data are considered, an extra complication
cannot be avoided, as detected signals suffer from both the finite 
resolution of antennae and pixilation effects. 
The total result is an effective reduction of the actual power
at a given multipole, and the measured APS become
\begin{equation}
C_{Y,l}^{meas} = C_{Y,l} |_sW_l|^2
\end{equation}
with
\begin{equation}
_sW_l =  b_{pix,l}\, _sB_l
\end{equation}
where $b_{pix,l}$ is the spherical harmonic decomposition of the pixel
window function, and $_sB_l$ is the beam window function. This writes
\begin{equation}
\label{windowf}
_sB_l =  \exp[ -(l(l+1)-s^2) \sigma^2 /2], 
\end{equation}
where $\sigma={\rm FWHM}/\sqrt{8\ln 2}$  for Gaussian
beams and $s=2$ for polarization, whereas
$s=0$ for anisotropy (Ng \& Liu 1999).
The rms polarization $P_{\mathrm{rms}}$ measured over a
finite--resolution map is then given by
\begin{equation}
P_{\mathrm{rms}}^2
= \sum_{l \geq 2} \frac{2l+1}{4\pi } C_{Pl}\, _2W_l^2,
\label{prms}
\end{equation}
where
\begin{equation}
C_{Pl}=C_{El}+C_{Bl}.
\end{equation}
When $Q$ and $U$ maps are available one can also use
$P_{\mathrm{rms}}^2=Q^2_{rms}+U^2_{rms}$).

As an example, let us consider an instrument with a FWHM aperture of $7^{\circ}$.
In this case the reduction caused by a factor $(_2W_l/b_{pix,l})^2$ is $e^2$
for $l\simeq 27$. It is then clear why, with such an aperture, it makes
sense to explore spherical harmonics up to $l=30$. Let us further recall
that, in order to obtain an optimal
exploitation of the signal,  it is convenient to set pixels
with centres at a distance $\sim \rm{FWHM}  /2$. 

\subsubsection{Small sky areas and non--Gaussian signals}
The treatment of the previous subsection can be simplified,
when a small patch of the sky is considered, so that the celestial
sphere can be treated as flat (Seljak 1997).
%[see e.g., Zaldarriaga (1998)]
Then the spin--weighted harmonic expansions (\ref{expansions}) reduce
to Fourier expansions in the $(\theta_x , \theta_y)$ plane:
\begin{equation}
\label{eq15}
Y(\vec{l})
= \Npix^{-1} \sum_k \exp(-i\vec{l}
                                \cdot \vec{\theta}_k)
Y(\vec{\theta}_k)~;
\end{equation}
here $\Npix$ is the number of pixels, and $Y$ stands
for either $Q$ or  $U$. Passing to $E$ and $B$ is also simpler, as
\begin{eqnarray}
E({\vec l}) &  =  &  Q({\vec l}) \cos(\phi_{\vec l}) +  U({\vec l}) \sin(\phi_{\vec l})  \nonumber \\ 
B({\vec l}) &  =  & -Q({\vec l}) \sin(\phi_{\vec l}) +  U({\vec l}) \cos(\phi_{\vec l})
\end{eqnarray}
$\phi_{\vec l}$ being the angle between ${\vec l}$ and the $\theta_x$ axis.
From the expansion coefficients, the APS can be recovered and read
\begin{equation}
\label{fouriercl}
C_{Yl} = \Omega \left\langle
Y    (\vec{l})
Y^{*}(\vec{l})
\right\rangle \mathbf{.}
\end{equation}
Here $\Omega $ is the size of the sky patch and the average $\langle\rangle$ is to be performed
 over those $\vec{l}$ having 
magnitude close to $l$. These spectral components,
referring to the portion of the sky we consider, are {\it local}
and can depend on the setting of such portion.

When dealing with foregrounds, which are non--Gaussian, APS can
still be useful to evaluate their contamination to the CMBR signal.
In such a case, an angle averaging should be explicitly
performed. This amounts to summing upon $m$ and dividing by $2l+1$.
Eqs.~(\ref{t1u2}) should not be used, because of the phase
coherence of harmonic amplitudes.
Furthermore, one should carefully distinguish between {\it local} and
{\it global} APS. Foreground subtraction ought to be performed
using local APS; in principle locality should be enhanced as much as
data uncertainties allow, in order to perform correct subtractions
on small angular scales.

It was not recognized for some time that, by expanding the polarized intensity
$\Pi =\sqrt{Q^2+U^2}$ in spherical harmonics, a further
 spectrum, $C_{\Pi l }$, is constructed. As first noted in 
Tucci \etal\ (2002)
$C_{\Pi l }$  conveys an information distinct from
$C_{Pl }$. In particular, $C_{\Pi l }$  provides an incomplete
information, since it does not include the polarization direction; 
for instance, it
cannot be used to build simulated polarization maps, and cannot
account for beam--depolarization effects. 
However, it is still useful when  studying the emission of Galactic
foregrounds at low frequency ($\leq 2$~GHz), where
$C_{El}$ and $C_{Bl}$ can be affected by Faraday
rotation, whereas
$C_{\Pi l}$ turns out to be quasi--independent of it
(Tucci \etal\ 2002), and can be safely extrapolated to
higher frequencies.

\subsection{Origins of anisotropies and polarization of the CMBR}
Photons undergoing Thomson scattering become polarized. In the presence
of uniform and isotropic electron and photon distributions, however,
angular averaging smears out polarization. To preserve some of it, not 
only must the photon distribution be anisotropic, but anisotropy must 
have a quadrupole component. Until photons and baryons are tightly 
coupled so as to form a single fluid, any quadrupole in the photon 
distribution is erased; accordingly, all scales entering horizon before 
recombination develop a quadrupole only after decoupling begins, so 
that photon and baryon inhomogeneities are out of phase.

When photons have acquired a mean--free--path of the order of 
fluctuation sizes, therefore, quadrupole anisotropies arise. Such 
a long mean--free--path, however, also means that the optical depth 
$\tau$ of baryonic materials is approaching unity (from above) and 
more and more photons are 
unscattered until present. 
However, only the restricted fraction of photons that still undergo a late
scattering, can acquire polarization. Therefore, primary polarization 
is small both because quadrupole is small and because few photons 
are scattered after it arises. Quadrupole (and, hence, polarization)
turns however higher on scales entering recombination with larger 
dipole, i.e. in the kinetic stage. The number of photons scattered 
decreases further for scales entering the horizon after recombination, 
where the polarization degree rapidly fades with scale increase.

All these effects can be followed in detail by numerical linear codes, like
CMBFAST\footnote{www.physics.nyu.edu/matiasz/CMBFAST/cmbfast.html},
able to compute the amount of polarization produced
on each scale as a function of the parameters of the underlying cosmological model.
The information that the tiny
polarization signal carries can therefore help to reduce any parameter
degeneracies that the stronger anisotropy signal cannot resolve.
In particular, the detailed history of recombination, occurring at
redshifts $z \sim 10^3$, can be inspected through these signals.

However,
it is easy
to see that the horizon scale at recombination occupies an angular
scale $\sim 10^{-2}$ radians, on the celestial sphere.
Accordingly, the primary polarization spectrum can be expected to peak
at $l \sim 10^2$--10$^3$ and to be negligible at $l < 30$, those $l$'s
we can inspect if ${\rm FWHM}\simeq 7^{\circ}$. The above
inspection is therefore precluded to the SPOrt experiment.

The polarization signal SPOrt aims at detecting must be related to later 
events, when the scales entering the horizon subtended angles 
$\sim 7^{\circ}$ 
or were even larger than so. Should the Universe be filled with neutral 
gas from recombination to now, no such polarization could be expected. 
It is known, on the contrary, that diffuse baryonic materials, today, 
are almost completely ionized. But this could not be enough: baryonic
materials are so diluted, today, that the scattering time for CMBR 
photons $t_s \simeq 4.45 \cdot 10^{18} \Omega_b^{-1} h^{-2}$s 
($h$: Hubble parameter in units of 100~km/s/Mpc) widely exceeds the 
Hubble time $t_H = 3.09 \cdot 10^{17} h^{-1}$s. Hence, {\it secondary} 
polarization can arise only if reionization is early enough not to be 
spoiled by baryon dilution. 

Let us therefore extrapolate these figures to the past: while $t_s 
\propto a^3$ ($a$: the scale factor), the dependence of $t_H$ on $a$ 
depends on the model. Numerical computations are avoided by assuming 
an expansion regime dominated by ordinary non--relativistic matter, 
as in a standard CDM model. Then an ionized Universe is opaque ($\tau=1$)
to CMBR 
at $z > z_{op} \simeq 5\, (\Omega_b h)^{-2/3}$. If $\Omega_b h^2 
\leq 0.022$, as from standard 
BBNS limits (see Dolgov 2002 for a recent review),
and $h \simeq 0.5$, we obtain $z_{op} > 64$. The redshift range 
where reionization may have occurred ($z_{ri}$) is safely below such 
$z_{op}$. A standard CDM model disagrees with data, but no reasonable 
cosmology allows a substantial reduction of $z_{op}$. A full 
re--scattering of CMBR at low redshift is therefore almost excluded, 
but the opacity $\tau$ can be non--neglegible if $z_{ri}$ is not lower
than $\sim 8$--10. In turn, this may cause secondary polarization.

\subsection{Relevance of secondary polarization}
\label{sec:tau}
Let us briefly remind which kind of physics is explored through
$\tau$ measurements and, hopefully, through a more sophisticated analysis,
leading to estimates of $z_{ri}$ and other related parameters.
They could shed new light on (i) the cosmic reionization
history and the physics of systems causing it; (ii) the expansion
laws during reionization and the cosmic {\it substance} ruling
the expansion.

Good reviews on point (i) were provided by Loeb \& Barkana (2001), 
Shapiro (2001), Maselli \etal\ (2003) and Haiman (2003). 
Most studies, assuming the flat $\Lambda$--dominated {\it concordance} 
model ($\Omega_m = 0.3$, $h=0.7$), until recently converged in setting 
$z_{ri} $ in the interval 8--10 and suggested that hydrogen ionized 
then and remained such until now. Reionization was attributed to 
radiation produced by metal--free star formation in early galaxies,
whose effects depend on several parameters, as the photon escape 
fraction $f_{esc}$ and the slope of the Salpeter Initial Mass 
Function (IMF). Among the limits 
that such theories must fulfil is that helium, at variance from hydrogen,
reionized much more recently. However, although leaving all parameters 
vary in the widest possible range, $z_{ri}$ values above 10 had not 
been envisaged. Other possible sources of reionization had been 
identified in AGN (Ricotti \& Ostriker 2003 and reference therein) 
as well as {\it exotic} sources like decaying heavy sterile neutrinos 
(see Hansen \& Haiman 2003).

These predictions were strengthened by Miralda--Escud\'e (2003)
finding that, if the ionizing flux observed today is extrapolated to 
high $z$, the photon flux is able to reionize the cosmic hydrogen at 
$z_{ri} \sim 8$--10 and $\tau \simeq 0.08$. Furthermore Djorgovski et 
al. (2001) and Becker \etal\ (2001) reported observations of QSO's at 
$z > 5$, showing an increase of the Ly$\alpha$ forest opacity when $z$ 
goes from 5 to 6. Becker \etal\ (2001)  also found 
Gunn--Peterson effect in one QSO at $z \simeq 6.3$. These data 
confirmed that a fraction of neutral hydrogen was present, in the
intergalactic medium, at $z \sim 6$. (It must be however reminded
that these very authors warned the reader against an immediate
conclusion that $z_{ri} \simeq 6$.) Let us also add that Bullock 
\etal\ (2000) revived the idea of the excess of galactic satellites, that 
N--body simulations predict, by considering the slowing down of 
baryon accretion on small velocity lumps after $z_{ri}$.
Their analysis apparently leads to requiring $z_{ri} < 10$,
at least for $\Lambda$CDM cosmologies.

\begin{figure}
\centering
\includegraphics[width=0.6\hsize]{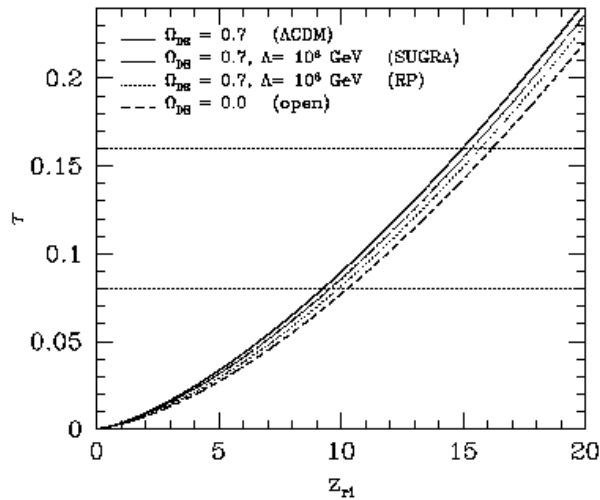}
\caption{Opacity $\tau$ as a function of the reionization redshift,
assuming instantaneous reionization at $z_{ri}$. Horizontal lines
refer to $\tau$ values  0.08 and 0.16. For $\tau > 0.16$ reionization
must have occurred at $z_{ri} > 15$, almost independently of the model.
}
\label{fig2}
\end{figure}

Assuming an (almost) immediate and complete reionization after $z_{ri}$,
in Fig.~\ref{fig2} we plot how $\tau$ depends on $z_{ri}$ itself, 
in various models, assuming $\Omega_b h^2 = 0.02$. A larger $\tau$ 
arises in flat models with low $\Omega_m$, but $\tau > 0.08$ can 
hardly be reconciled with $z_{ri} < 10$ while, for the 
aforementioned range of $z_{ri}$, the range of $\tau$ values lies typically 
below 0.05.

Such a satisfactory self--consistent picture has been put in
severe danger by the
recent WMAP release. Using their first--year observation of $C_{TEl}$
at low $l$ values, Kogut \etal\ (2003) estimate that, independently
from any model assumption, $\tau = 0.17 \pm 0.04$. However, Spergel
\etal\ (2003) comment that the low value of the quadrupole anisotropy
and, in general, the rather low values of $C_{Tl}$ at $l < 10$
disfavour very high $\tau$ value, as reionization should enhance large
angular scales anisotropies (see, however, Efstathiou 2003). %%%SB
In fact, once a concordance model is
assumed, the likelihood distribution is quite flat towards lower $\tau$
values and decreases just of 0.05 as $\tau$ shifts from 0.19 to 0.11.
%%%%%%%%%%%%%%%%%%%%%%%%%%%%%%%%%
%%%%%%%%%%%%%%%%%%%%%%%%%%%%%%%%%
%%%%%%%%%%%%%%%%%%%%%%%%%%%%%%%%%
\begin{figure} [htp]
\centering
\includegraphics[width=0.7\hsize,clip]{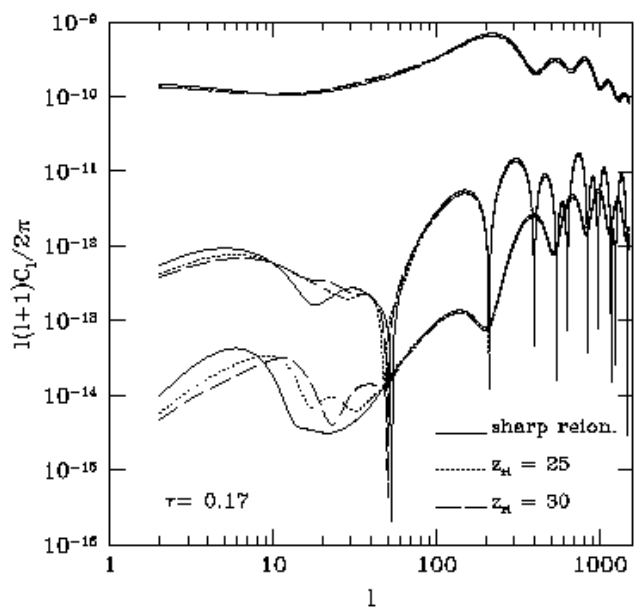}
\caption{Dependence of angular power spectra (in K$^2$) on model 
reionization histories.}
\label{new1}
\end{figure}
%%%%%%%%%%%%%%%%%%%%%%%%%%%%%%%%%

Bearing in mind all these reserves and also the fact that, at
the 3--$\sigma$ level, $\tau \sim 0.05$ is still consistent with
observations, if we leave apart any theoretical bias, we must
also accept that, e.g., a value of $\tau \sim 0.25$ is no longer
excluded by the best observations now avaliable.

Such findings immediately reopened the theoretical discussion
on the reionization physics. Values of $\tau \sim 0.17$ were soon
found to be compatible with galaxy formation simulations by 
Ciardi \etal\ (2003), with reionization caused by stars in
galaxies with total masses $\sim 10^9$--$10^{10} M_\odot$, assuming
a (moderately) top--heavy Salpeter IMF, a high stellar production
of ionizing photons, and a fairly high 
$f_{esc} \simeq 0.05$. These values are however hardly compatible 
with Gunn--Peterson effect findings.

A more speculative perspective becomes however more realistic
in the light of high $\tau$ values: reionization might have
occurred twice. Models where this occurred have been proposed by
Cen (2003), Wyithe \etal\ (2003) and Sokasian \etal\ (2003). A redshift 
interval $\Delta z 
\simeq 3$, when $H$ is only partially ionized, should have elapsed
between the two reionization stages. Temporary recombination might be
due to feedback by SNe reducing star formation efficiency (see,
e.g., Scannapieco \etal\ 2000, Madau \etal\ 2001); otherwise,
a similar SNe action, concentrated in suitable spatial volumes,
caused a partial temporary recombination in regions where QSO with
Gunn--Peterson effect have been seen.
%%%%%%%%%%%%%%%%%%%%%%%%%%%%%%%%%
\begin{figure} [htp]
\centering
\vskip -1.truecm
\includegraphics[width=0.7\hsize,clip]{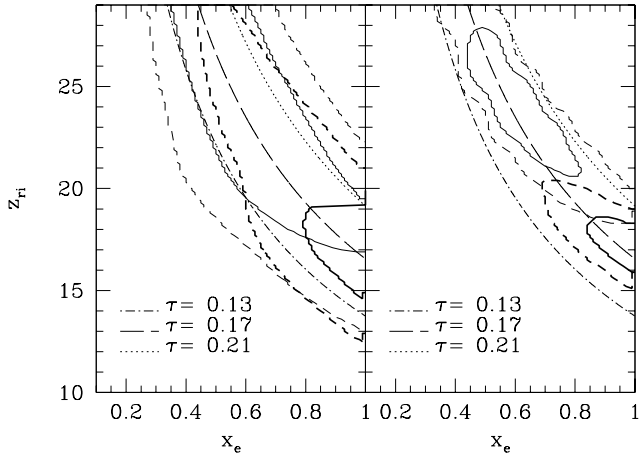}
\vskip -1.truecm
\caption{Likelihood distributions for the simultaneous determination
of reionization redshift $z_{ri}$ and ionization rate $x_e$ in models
with fixed $\tau = 0.17$; the left (right) panel corresponds to
a $7^{\circ}$--pixel noise $\Spix = 2\, \mu$K (0.5$\, \mu$K).
}
\label{new2}
\end{figure}
%%%%%%%%%%%%%%%%%%%%%%%%%%%%%%%%%

Within this context, let us remind that cosmological
models with equal $\tau$, but different reionization histories,
predict different CMBR anisotropy and polarization. This is shown,
first of all, in Fig.~\ref{new1}, where APS, for a fixed $\tau$ value,
but different reionization histories, are shown. The other parameters 
of this spatially flat model are $\Omega_b h^2= 0.02$, $\Omega_m=0.3$, 
$h = 0.7$. 

Besides the
case of a sharp reionization, taking place at the nearest possible
redshift, we also consider reionizations starting at larger $z_{ri}$,
followed by a single zero--ionization interval, suitably set to recover the
fixed $\tau$ value. Notice, in particular, that the high--$l$,
apparently, does not depend on $z_{ri}$, while the shift, at low
$l$, is quite significant.

The plot in Fig.~\ref{new1} was made with a suitable generalization
of the CMBFAST program. Using the techniques described in
Sect.~\ref{likelihood}, we can 
estimate likelihood distributions
and see that the differences shown in Fig.~\ref{new1} 
enable us to set significant constraints on cosmological parameters,
already with a sensitivity close to SPOrt's.

In Fig.~\ref{new2} 
(%taken from Colombo 2003;
see 
%also 
Colombo \etal\ 2003; Mainini \etal\ 2003), we consider two models 
with $\tau=0.17$,
either due to full reionization ($x_e = 1.0$) arising at the nearest possible
redshift ($\simeq 17$) or to a reionization with constant $x_e = 0.6$,
again since the nearest possible
redshift ($\simeq 24$); the other model parameters are as
in Fig.~\ref{new1}.

In both plots
1-- and 2--$\sigma$ confidence curves for the simultaneous 
determination of $x_e$ and $z_{ri}$ are shown, by solid
and dashed lines, respectively. Such curves are obtained by
averaging over 1000 realizations of each model.
The plots are also crossed by constant--$\tau$ curves corresponding
to three $\tau$ values. The l.h.s. (r.h.s.) plot corresponds to a pixel
noise $\Spix = 2\, \mu$K (0.5$\, \mu$K), where the pixel size is about $7^{\circ}$. 

A first observation is that $\tau$ is more easily
determined than $x_e$ and $z_{ri}$, separately. 
Furthermore,
it is clear that the level of pixel noise that SPOrt is likely
to achieve ($\Spix \simeq  2\, \mu$K) is just 
marginal to distinguish between the two
models. 
 If $\Spix = 0.5\, \mu$K, 
instead, at the 1--$\sigma$ level, confidence curves are
disconnected and have just a marginal intersection at the
2--$\sigma$ level. 
The capability of SPOrt to distinguish reionization
histories cannot be excluded and depends on the model realization in 
the real Universe.

In principle, however, the aforementioned figures show that an 
analysis of CMBR 
polarization data is an excellent probe into the reionization pattern of the 
Universe and, 
therefore, on birth, evolution and death of early starlike objects. 

\begin{figure} [htp]
\centering
\vskip -3.3truecm
\includegraphics[width=0.8\hsize,clip]{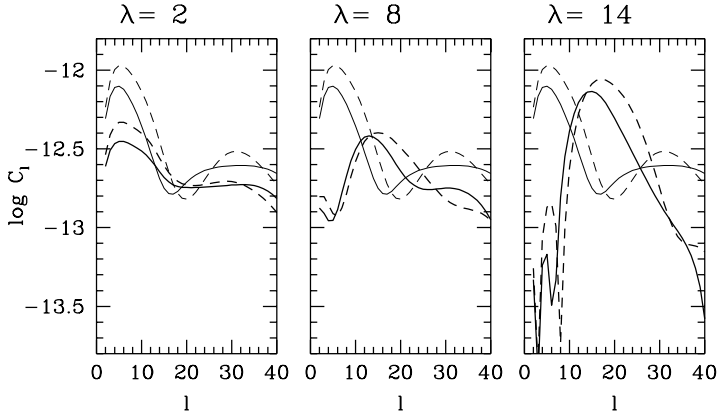}
\vskip -2.1truecm
\caption{TE correlation spectra (in K$^2$) for 
$\Lambda$CDM and RP models compared.
Here $\lambda = \log(\Lambda$/GeV). Solid and dashed lines refer to
$\tau=0.14$ and $\tau=0.20$, respectively. 
}
\label{new3}
\end{figure}

Let us now come to the point (ii): in shaping polarization spectra, 
the nature of the Dark Energy (DE) component of the Universe plays a role. 
For the sake of example, in Fig.~\ref{new3} 
we show the $l$--dependence of $C_{TEl}$ 
for $l$ up to 40. In all plots $\Omega_b = 0.05$, $h=0.7$, 
$\Omega_m=0.3$, while the solid (dashed) curve refers to $\tau = 0.14$
(0.20). In all plots the spectra of a $\Lambda$CDM model are shown,
together with the spectra for models where DE is due to a scalar
field $\phi$, self--interacting through a Ratra--Peebles (1988,
RP hereafter) 
power--law potential $V(\phi) = \Lambda^{\alpha+4}/\phi^{\alpha}$.
[In some figure of this paper we consider also SUGRA
potentials $V(\phi) = (\Lambda^{\alpha+4}/\phi^{\alpha})\exp (4\pi G\phi^2)$,
Brax \& Martins 1999, 2000].
The three plots differ for the choice of the energy scale $\Lambda$
and $\lambda = \log_{10}(\Lambda/{\rm GeV})$. The differences
between $\Lambda$CDM and RP models, already significant for $\lambda
= 2$, become huge for $\lambda = 14$, where the first correlation
peak has even negative values.

\begin{figure} [htp]
\centering
\includegraphics[width=0.9\hsize]{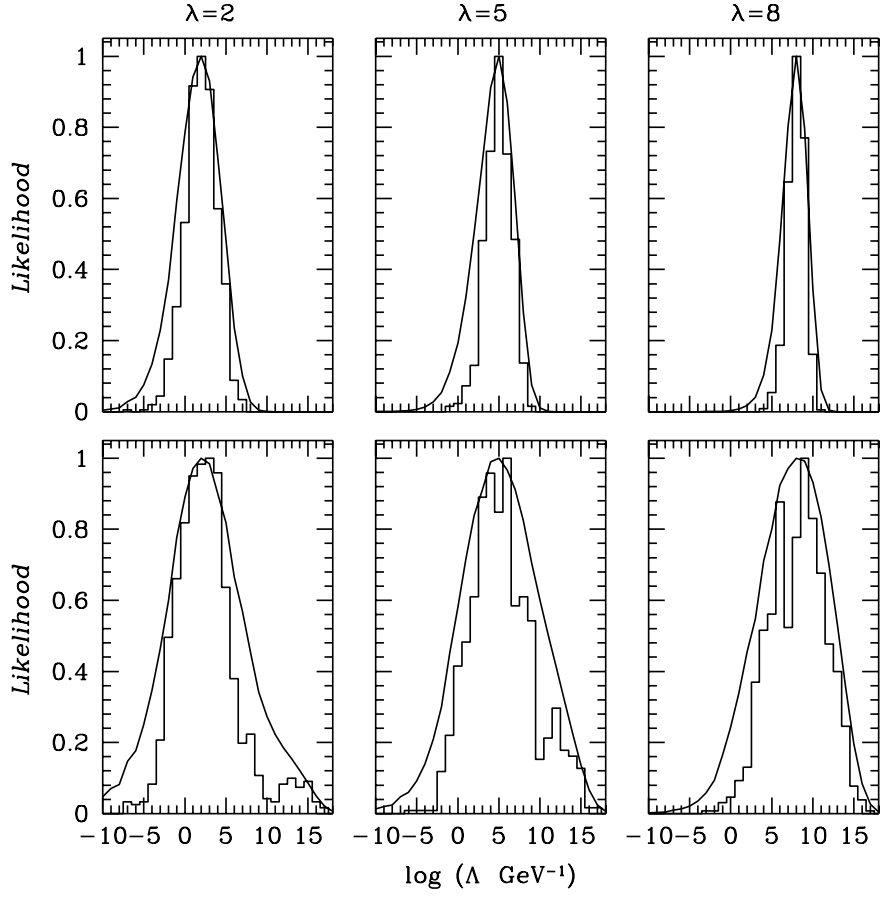}
\caption{Likelihood distribution on $\Lambda$ (solid line), averaged on 
cosmic and noise variance, in measures of artificial data, built from 
different RP models ($\lambda = \log(\Lambda$/GeV)).  
Upper (lower) plots refer to a pixel noise of 0.2$\, \mu$K  (2.0$\, 
\mu$K). Histograms yield the distribution of the likelihood peak in the 
1000 realizations considered.
}
\label{new4}
\end{figure}

The impact of DE nature can be even better appreciated in Fig.~\ref{new4},
where we plot the likelihood distributions on $\Lambda$, as obtainable
when artificial data built from RP models, with the $\lambda$ scale
indicated in top of each frame, are ``observed''. Here again, the 
techniques described in Sect. \ref{likelihood} have been used. In the 
plots, the histograms report the distribution of the peak likelihood,
as obtained from 1000 realizations of each model. The solid curves,
instead, are the likelihood distributions, as obtained by
averaging over all realization. Upper (lower) plots refer to a pixel
noise of 0.2$\, \mu$K (2.0$\, \mu$K). 

While, for the unlikely model $\lambda = 8$, the RP potential is
discriminated from $\Lambda$CDM at more than 3--$\sigma$'s,
at the noise level expected for
SPOrt, the discrimination,
with such noise and for lower $\lambda$'s, is not safer than 1 or 2 
$\sigma$'s. These plots, however, put in evidence that, even for
$\sigma = 0.2\, \mu$K, the histogram of the peak distribution lays
well inside the average likelihood distribution, so outlining
that we are still far from a regime where cosmic variance dominates.

%%%%%%%%%%%%%%%%%%%%%%%%%%%%%%%%%
%%%%%%%%%%%%%%%%%%%%%%%%%%%%%%%%%
%%%%%%%%%%%%%%%%%%%%%%%%%%%%%%%%%
%%%%%%%%%%%%%%%%%%%%%%%%%%%%%%%%%

Besides considering the point (i) and (ii) separately, one
should discuss how far DE parameters and recombination history
can be simultaneously studied, at different sensitivity levels.
This is an open question that will be debated elsewhere.

All these arguments converge to indicate the importance
of inspecting polarization at low $l$, where we are still quite
far from meeting cosmic variance limits.

%%%%%%%%%%%%%%%%%%%%%%%%%%%%%%%%%%%%%%%%%%%%%%%%%%%%%%%%%%%%%%%%%%%%%%%%

\subsection{The foregrounds}
Detecting a cosmological signal requires a fair subtraction of all
foreground signals. Depending on frequency,
this may be not so simple; foregrounds, however,
can be separated from cosmological backgrounds because of their
spectral properties and their non--Gaussian nature.

Galactic foregrounds, e.g., have strongly non--Gaussian distributions,
so that their APS do not provide a complete statistical description.
The coordinate space description is often more satisfactory. As a
matter of fact, the angular distribution of total intensity foregrounds
has been provided by WMAP, at five frequencies between 23 and 94 GHz.
The published maps (Bennett \etal\ 2003) have a $1^{\circ }$ resolution;
around this scale, it is recognized that CMBR anisotropies dominate
in total intensity for Galactic latitudes $|b|>15^{\circ }$ and
frequencies $\nu =30$--$100\, $GHz.

However, the APS are very convenient for storage of information with a clear
separation of angular scales, and by analogy with previous CMBR studies,
they have been applied to foregrounds for about a decade, especially for
extensive predictions at many scales. Such investigations must be performed
at many frequencies since the APS of any backgrounds except CMBR depend both
on $l$ and $\nu $. A common simplifying assumption is
\begin{equation}
C_{Yl}=A_Y(\nu )l^{-\alpha _Y}  \label{forespec}
\end{equation}
with $\alpha _Y$ independent of both $\nu $ and $l$. This is similar to the
common real--space assumption, that the frequency spectral index be
independent of sky position, and is likely to be as much unsatisfactory for
an accurate description of data.

Until 2000 the only available experimental knowledge regarded total
intensity, and rather steep APS were often supported for Galactic
foregrounds. For dust emission $\alpha _T^{\mathrm{(D)}}\simeq 3$ was found
from IRAS (Gautier \etal\ 1992) and DIRBE (Wright 1998); from the combined
DIRBE and IRAS maps a slightly flatter slope, $\alpha _T^{\mathrm{(D)}%
}\simeq 2.5$, was derived by Schlegel \etal\ (1998). For free--free emission,
correlating COBE--DMR with DIRBE gave $\alpha _T^{\mathrm{(FF)}}\simeq 3$
(Kogut \etal\ 1996). However modelling free--free emission from H$\alpha $
maps leads to a flatter slope $\alpha _T^{\mathrm{(FF)}}\simeq 2.3$ and a
much lower normalization at 53 GHz (Veeravagham \& Davies 1997). This
discrepancy supported the case for anomalous dust emission in the microwave
region down to a few GHz, and Draine \& Lazarian (1998,1999) proposed two
possible mechanisms for it (rotational excitation of small grains, and
magnetic grains). Some further evidence for anomalous emission has been
proposed over the years [see, e.g. Lazarian \& Prunet (2002) for a short
review]; however, extensive investigations of available low--latitude
surveys, carried out within the SPOrt project (Tucci \etal\ 2000, 2001,
 2002),
supported a more conventional view, that the total intensity APS around a
few GHz are generally dominated by synchrotron, except in low emission
regions where unresolved point sources dominate especially for 
$l\ga 10^2$
(cfr. Sect. \ref{synchaps}). WMAP's analysis shows that the contribution of
spinning dust, if any, is less than 5\% the foreground intensity at 23 GHz,
and this casts serious doubts on the relevance of such emission at any
frequency. The correlation of radio and dust emissions is simply explained
as due to the fact that both trace star--formation activity. WMAP's total
foreground APS exhibit moderate slopes, $\alpha _T\simeq 2$ for 
$l\la 10^2$
in all frequency bands. Since thermal dust dominates at 94 GHz, we can
easily infer $\alpha _T^{\mathrm{(D)}}\simeq 2$; the same conclusion is
likely to apply to free--free and synchrotron which appear to be mixed at
23--41 GHz.

For the synchrotron total intensity, results on the slope of the
low--frequency angular spectrum have been published since 1996, based on
several surveys such as the 408~MHz maps of Haslam \etal\ (1981), the
1420~MHz northern--sky survey of Reich \& Reich (1986) and the 2326~MHz
Rhodes/HartRAO survey (Jonas \etal\ 1998). Slopes $\alpha _T^{\mathrm{(S)}%
}\simeq 2.5\div 3$ were given by Tegmark \& Efstathiou (1996), Bouchet et
al. (1996) and Bouchet \& Gispert (1999); for the Rhodes/HartRAO survey
Giardino \etal\ (2001) report a Galactic latitude dependence, with $\alpha
_T^{\mathrm{(S)}}\simeq 2.4$ in the full maps and $\alpha _T^{\mathrm{(S)}%
}\simeq 2.9$ at $\left| b\right| >20^{\circ }$ down to the resolution limit $%
\sim 1^{\circ }$. However, at the same frequencies an analysis of the
Tenerife patch reported by Lasenby (1997) provided a nearly scale--invariant
spectrum $\alpha _T^{\mathrm{(S)}}\simeq 2$; see also the analysis of the
5~GHz Jodrell Bank interferometer by Giardino \etal\ (2000). More recently,
local APS slopes in the Galactic Plane at 2.4 and 2.7~GHz were found to be 
$\alpha _T^{\mathrm{(S)}}\simeq 1.7$ (Tucci \etal\ 2000; Bruscoli \etal\ 2002).
 This
result is expected to be affected by free--free emitting point sources.
However, as already remarked, the WMAP full--sky maps give moderate slopes
for the APS in all of its frequency bands; therefore the low slopes found 
by Tucci \etal\ (2000) are confirmed.
We can infer $\alpha _T^{\mathrm{%
(S)}}\simeq 2$, although a detailed analysis of APS of the individual
foreground components has not been provided yet.

Clearly, foregrounds with APS slopes $\alpha _T\sim 2$ or larger are
relatively more important at large angular scales. An important exception is
given by unresolved point sources, for which flat intensity APS are
predicted, $\alpha _T^{\mathrm{(PS)}}=0,$ provided their distribution in
unstructured (shot noise). Modelling for populations of extragalactic radio
sources has provided predictions for the amplitude factor $A_Y(\nu )$
(Toffolatti \etal\ 1998; Guiderdoni 1999); however Galactic HII regions also
contribute, and should dominate near the Galactic Plane. Point sources must
be the most important foreground at sufficiently high $l$; for instance, for
the BOOMERANG 153~GHz experiment this happens at the smallest angular scales
of the measurement, i.e. $l\simeq 10^3$ (Netterfield \etal\ 2002). WMAP's
APS  for $l \ge 200$ at 41, 61 and 94 GHz exhibit a similar behaviour,
which however may be due to instrumental noise.

\subsubsection{Dust polarization APS}

While waiting for a more complete analysis of the WMAP data, at 
the time of writing
direct measurements of foreground $Q$ and $U$ fields, or $E$ and $B$ APS, are
still lacking. Studies of Galactic polarization APS have been initiated by
the work of Prunet \etal\ (1998), who modelled polarized cross--sections
assigning grain compositions and shapes, the distribution of Galactic dust
using the HI Dwingaloo survey, and the magnetic field by means of gas
density structures (assuming the field lines to be parallel or perpendicular
to gas filaments). A template for polarized emission from thermal dust was
obtained, and applying the discrete Fourier transform to it in the
small--scale limit, local APS were derived for 
$l\la2000$. 
The resulting
slopes are $\alpha _E\simeq 1.3$, $\alpha _B\simeq 1.4$ and $\alpha
_{TE}\simeq 1.95$, and predictions are provided for the scenario to be met
by PLANCK at 142--217 GHz. Although some assumptions underlying the model
are now regarded as simplistic by Lazarian \& Prunet (2002), the importance
of this paper should not be underrated, because it clearly recognized that
polarization APS cannot simply mimic intensity APS up to a constant
polarization degree. In a paper studying the  
$C_{\Pi l}$ 
spectrum of
starlight (i.e., of the distribution of star polarization vectors over the
sky), Fosalba \etal\ (2002) notice that polarized 
emission in the sub--mm/FIR
should be related to optical polarization by differential absorption, and
suggest the use of starlight data to improve modelling of polarized dust
emission.

In the light of WMAP and previous results, we do not expect spinning dust to
be important in polarization. Interesting effects however might arise at
some frequencies in the magnetic--grain scenario. For grain models with
random Fe inclusions the polarized emission appears to be peaked around $\nu
\approx 80$ GHz, and differing from the usual expectation for thermal
emission (but also for spinning dust), the polarization vector should be
parallel to the Galactic magnetic field, as is the case for starlight
polarization by differential absorption. APS modelling has not been provided
so far for the anomalous dust emission; the best estimates for spinning dust
provided by Tegmark \etal\ (2000) simply assume the same APS behaviour as
for  thermal dust.

\subsubsection{Galactic synchrotron APS\label{synchaps}}

Differing from total emission, polarized Galactic emission is not expected
to be much affected by free--free in the microwave region. Most
investigations have thereby been focused on synchrotron. Starting with the
work of Tucci \etal\ (2000), a number of papers tackled the problem of
polarized synchrotron APS at low frequencies in view of extrapolations to
the cosmological window. In particular, within the ambit of the SPOrt
project the following surveys were analysed:

\begin{itemize}
\item  The Parkes 2.417 GHz survey of the southern Galactic Plane, described
by Duncan \etal\ (1997) (henceforth, D97). It covers a strip, $127^{\circ }$
long, with a FWHM resolution of  
$10.4^{\prime }$.

\item  The Effelsberg 2.695 GHz survey (Duncan \etal\ 1999: D99), covering a
$70^{\circ }$ strip of the first--quadrant Galactic Plane with a resolution
of $5.1^{\prime }$.

\item  The Effelsberg 1.4 survey (Uyaniker \etal\ 1999: U99) with five
disjoint regions fairly close to the Galactic Plane, $-15^{\circ }\leq b\leq
20^{\circ }$. The angular resolution is there $9.35^{\prime }$.

\item  The Leiden survey of Brouw \& Spoelstra (1976) (BS76). It covers
about 40\% of the sky up to near the north Galactic Pole at five
frequencies, between 408 and 1411 MHz, with resolution varying between $%
2.3^{\circ }$ and $0.6^{\circ }$, but with a significant undersampling
(BS76).

\item  The ATCA interferometric survey of the SGPS Test Region, a $4^{\circ
}\times 7^{\circ }$ region internal to the Parkes survey, with an elliptic
beam of $67^{\prime \prime }\times 87^{\prime \prime }$ (Gaensler \etal\
2001: G01). Only the 1.404 GHz channel has been used so far for computation
of APS.
\end{itemize}

%%%%%%%
\begin{figure}
\centering
\includegraphics[width=0.9\hsize]{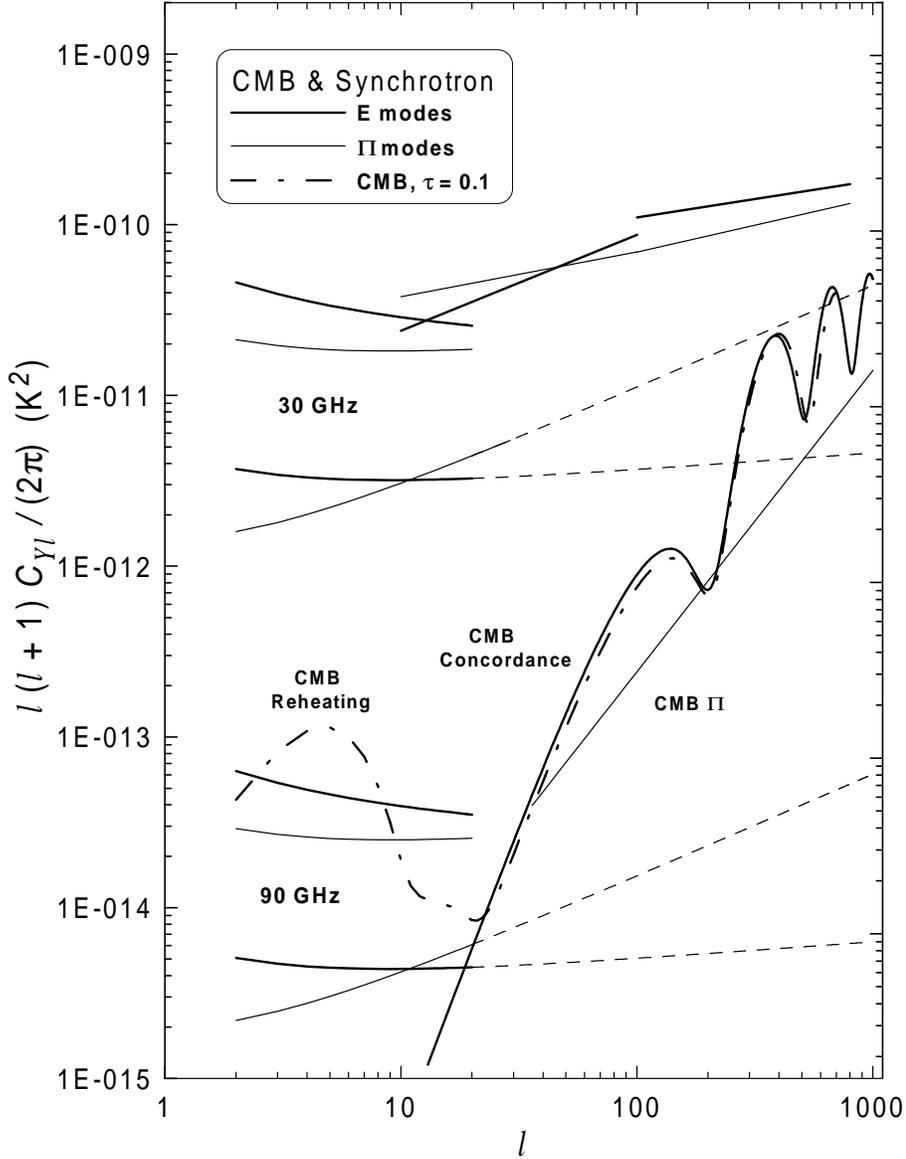}
\caption
{The APS for $E$ mode (thick and dash--dotted lines) and polarized intensity
(thin lines). The CMBR concordance model assumes the parameters adopted by
Kovac \etal\ (2002). The thin and dash--dotted CMB lines refer to
a model with a reheating optical depth $\tau =0.1$; the polarized intensity
curve is a power--law approximation to the output of simulations reported by
Tucci \etal\ (2002). The synchrotron APS are best--fit curves,
extrapolated to 30 and 90 GHz from the following maps: The template of
Bernardi \etal\ (2003), $ l=2$--20;
olation; 
the intermediate--latitude patch in BS76 (Bruscoli \etal\ 2002),
$l=10$--70; the D97 survey (Tucci \etal\ 2000), $l=100$--800.
For the template of Bernardi et al. (2003) we also give the APS from 
reduced maps at high Galactic latitudes, with extrapolations
to  $ l=20$--$1000$ represented by dotted lines. }
\label{cmbsynch}
\end{figure}
%%%%%%%

The D97 and D99 surveys showed that Galactic--Plane polarization is not much
correlated to total intensity, being often high in the absence of sources in
total intensity, and exhibits a foreground component generally supposed to
originate locally, i.e. within a few kpc. Local APS have been computed in
twelve D97 patches and six D99 patches of size $10^{\circ }\times 10^{\circ
} $ (Tucci \etal\ 2000, 2001) and exhibit moderate slopes $\alpha _E$ $\simeq
\alpha _B\simeq 1.5$, with no systematic differences between high-- and
low--polarized emission regions. A recent re--analysis of D97 data by another
group (Giardino \etal\ 2002) gives similar results. Out of the Galactic
plane, the U99 survey shows larger dispersions in the spectral parameters,
due to structures with strong polarized signals (e.g., in the Cygnus and Fan
regions) and regions of very low total intensity. Five low--intensity patches
with $\alpha _T$ $\simeq 0$, which must be dominated in intensity by
discrete sources, have been used to infer a rather accurate estimate 
$C_{Tl}^{\mathrm{(PS)}}\simeq 5\times 10^{-7}$ K$^2$ for the sum of all point
source populations at 1.4 GHz. Adopting a radio--source polarization degree $%
\sim 5\%$, the estimate 
$C_{Pl}^{\mathrm{(PS)}}\sim 1.3\times 10^{-9}$ K$^2$
was provided at 1.4 GHz, as well as other estimates at
different frequencies with reasonable assumptions on the frequency spectral
index (Bruscoli \etal\ 2002). Such estimates imply that the contribution of
point sources should be negligible for all of the \textit{polarization} APS
reported for all of the aforementioned surveys, as 
confirmed by the fact, that the
slopes of polarization APS do not show any flattening towards higher $l$.
The impact of Faraday rotation, however, is not clear from the data, and
must be investigated looking at different frequencies and angular scales. A
multiwavelength study has been performed in the range $l\leq 70$ by Bruscoli
\etal\ (2002) using three large patches of the BS76 survey at
various latitudes up to near the Galactic Pole. At such scales no definite
trends were found in the spectral slopes for increasing Galactic latitude. A
clear flattening of polarization APS, $\alpha _{E,B}\simeq 1$, appears at
low frequencies, $\nu \leq 610$ MHz, where Faraday rotation is larger; on
the other hand, at 1.4 GHz no significant dependence on Galactic latitude
appears, due to the large error bars.
Also
the APS normalizations are essentially consistent, as shown by 
Fig.~\ref{cmbsynch}. According to the analysis of D97, D99 and BS76 data, the
$C_{\Pi %
l}$ spectra at $\nu \geq 1.4$ GHz may be only slightly steeper than
$C_{Pl}$%
, since the best slope estimate is $\alpha _\Pi \simeq 1.6\div1.7$ and the
difference has little statistical significance. The substantial agreement
does not prove, however, that the polarization APS at $\nu \geq 1.4$ GHz do
reflect the intrinsic properties of synchrotron at all scales.

A more recent analysis of Tucci \etal\ (2002) 
in fact shows that this is not
the case at 1411 MHz at high resolutions. Rich small--scale structures appear
in the ATCA maps (G01) and other polarimetric observations. The sharp--edge
patches in the G01 maps of polarization angles are particularly impressive,
and strongly suggest that many structures are in fact produced by 
sharp Faraday--rotation changes in the polarization
angle. This has a strong bearing on $C_{El }$ and
$C_{Bl }$: The
polarization APS computed in a $4^{\circ }\times 4^{\circ }$ box and several
smaller patches give $\alpha _E$ $\simeq \alpha _B\simeq 2.8,$ whereas $%
\alpha _\Pi {}\simeq 1.6$. Such values have been compared to those computed
in an overlapping $5^{\circ }\times 5^{\circ }$ box extracted from the D97
survey. Large discrepancies are found for $\alpha _E$ and $\alpha _B,$ $%
\Delta \alpha \simeq 1$, but not for $\alpha _\Pi $. Rescaling the data with
a synchrotron spectral index $\beta _{\mathrm{sync}}$ between $-2.5$ and $%
-3, $ The ATCA $C_{\Pi l }$ spectrum appears to be the natural extension
to
smaller scales of the corresponding spectrum at $l \leq 800$, since the
normalization is also quite consistent. On the other hand, the power excess
of $E$ and $B$ modes in the ATCA spectra is nearly one order of magnitude at
scales $600\la l \la 1000$ 
and declines at $l$ greater than a few thousand.
This excess must be due to Faraday boosts of the polarization 
angles (Tucci \etal\ 2002).

\subsubsection{A synchrotron template at SPOrt scales\label{templat}}
In the light of the aforementioned results, a simple extrapolation of
$C_{El}$ and $C_{Bl}$ 
from $\nu \sim 1$ GHz to higher frequencies is not possible using an
$l$--independent spectral index. How can we get rid of Faraday rotation?
Giardino \etal\ (2002) use D97, the total--intensity survey of Haslam \etal\
(1981) and other radio data to produce a real--space model of polarized
synchrotron emission $T$ and $\Pi $; the polarization angle however is just
taken as a random quantity, so the final synthetic maps were declared to be
a toy model. A more powerful approach has been recently introduced by
Bernardi \etal\ (2003), intended to produce a Faraday--free template of
polarized synchrotron. It is based on the following steps: (a) A synchrotron
intensity map is built using the 408 MHz total intensity survey of 
Haslam \etal\ (1981) and the 1.4 GHz survey of Reich \& Reich (1986). 
Two different
frequencies are necessary to clean the maps from free--free emission. (b) The
polarized intensity $\Pi $ is modelled, assuming that polarization is
produced within a spherical ``polarization horizon'' and calibrating a
normalization factor with comparison with BS76 in the Fan region. (c) A map
of polarization angles is built from starlight polarization data, with
interpolation from Heiles (2000) catalog and a $90^{\circ }$ rotation. The
last step, which assumes an essentially common location for dust selective
absorption of starlight and for synchrotron emission, is of crucial
importance. Several tests discussed by Bernardi \etal\ (2003) support the
reliability of the template under this respect, and in connection with the
polarization horizon modelling for $\Pi $ as well. Due to the Heiles data
sampling the template has been provided for an angular scale of $7^{\circ }$%
. An update of the template, taking advantage of the WMAP intensity maps at
23 GHz, is in preparation.
APS computed on Faraday--free templates can be extrapolated to higher
frequencies better than those of current surveys. Due to the low resolution,
results could be provided by Bernardi \etal\ (2003) only for $l\leq
20$, where the full--map template gives  
$\alpha _P=1.9\pm 0.2$ and $\alpha _\Pi =1.9\pm 0.2$. The
substantial agreement with surveys at $2.4$--$2.7$ GHz reinforces both
the validity of the template and our previous support 
to moderate spectral slopes.
The high slopes $\alpha _{E,B}\simeq 2.8$ found
by ATCA in the SGPS Test Region at
1.4 GHz, on the other hand, must overrate intrinsic synchrotron slopes
substantially, being due to the high--$l$ decline of the spectral power
excess.

In Fig.~\ref{cmbsynch} we have collected the extrapolations to 30 and 90 GHz
of some of the power--law fits to $E$--mode and polarized intensity APS,
without attempting to give error bars. Such extrapolations, which at present
provide only reasonable guidelines,
cover the entire range $l=2$--$1000$
including results from various surveys at $\nu =1.4\div 2.4$ GHz and the 
aforementioned
template. For the template we show the APS derived from both  full maps and
reduced maps (50\%) containing only the lowest--signal pixels; for the latter,
the extrapolation to the angular range
$l=20\div 1000$ is also provided. Clearly, in view of the differences 
in the investigated sky
regions, coverages and frequencies, one should not expect to find a detailed
agreement. However,  the curves displayed in Fig.~\ref{cmbsynch} 
are roughly consistent with each other even for the normalization factor. 
Only the template APS derived from the reduced maps 
(with low--signal pixels) have a lower normalization,
and suggest that the detection of the secondary--ionization spectral bump should
not be hampered by synchrotron at 90 GHz.

\section{The SPOrt Experiment}
\label{sport}

The SPOrt experiment, carried on under the scientific responsibility 
of an international collaboration of Institutes headed by the IASF--CNR in 
Bologna, and fully funded by the Italian Space Agency (ASI),
was selected by ESA in 1997 to be flown on board the International 
Space Station (ISS) during the Early Utilization Phase. 

\begin{figure*}
\begin{center}
\includegraphics[width=0.46\hsize]{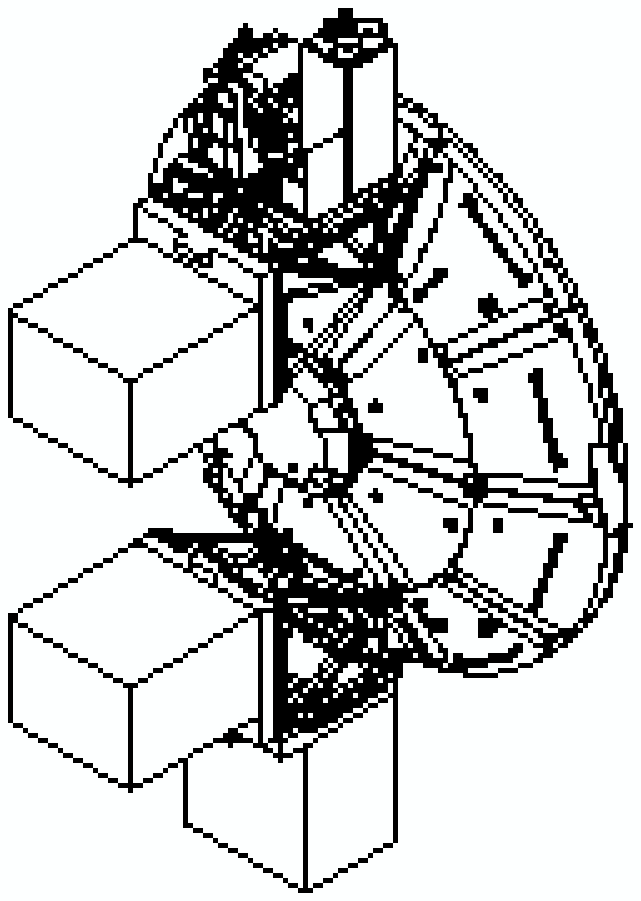}
\includegraphics[width=0.28\hsize]{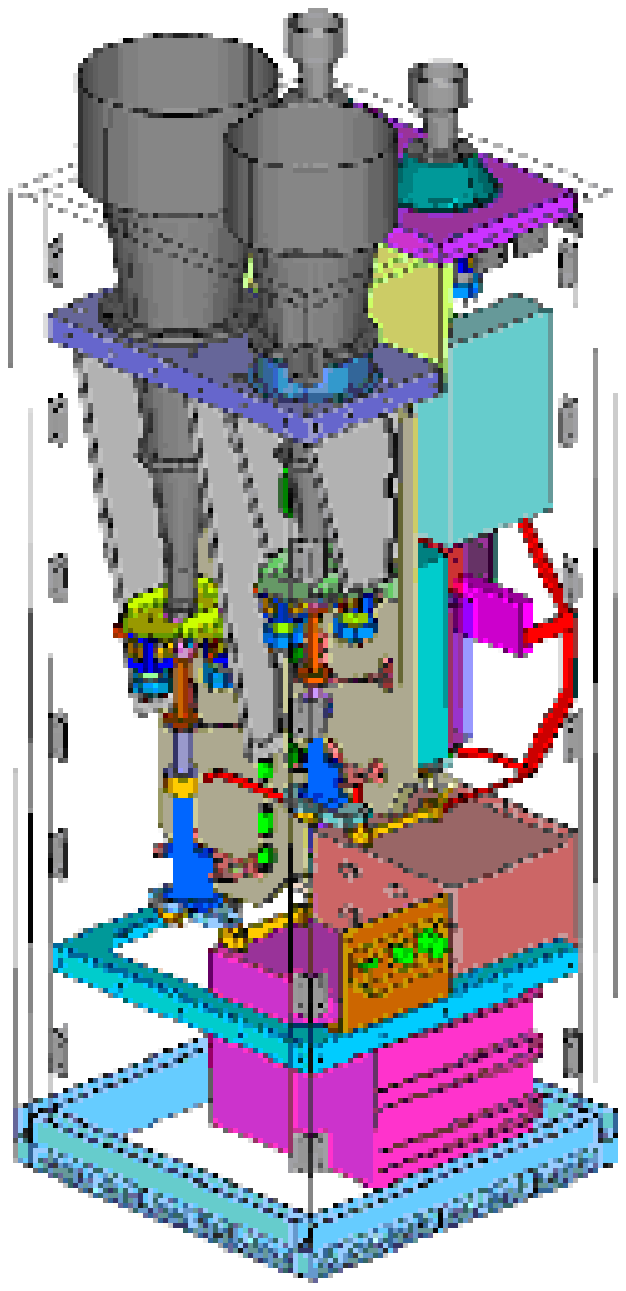}
\end{center}
\caption{Left: The SPOrt position on the Columbus 
External Payload Facility, onboard the ISS (courtesy by Alenia Spazio).
Right: Expansion of the SPOrt payload.}
\label{columbusfig}
\end{figure*}

The payload, shown in Fig.~\ref{columbusfig}, houses
four corrugated feed horns feeding a 22, a 32, and a pair of 90~GHz channels
providing direct measurements of the $Q$ and $U$ Stokes parameters.
It will be located on the Columbus External Payload Facility 
of the ISS in 2006, for a mission with a minimum lifetime of 18 months.

%%%%%%%%%%%%%%%%%%%%%%%%%%%%%%%%%%
%%%%%%%%%%%%%%%%%%%%%%%%%%%%%%%%%%%%%
%%%%%%%%%%%%%%%%%%%%%%%%%%%%%%%%%%%
\begin{table}
\caption{SPOrt main features: $\Npix$ is the number of FWHM
pixels covered by SPOrt,
$\sigma_{1s}$
is the istantaneous sensitivity (1 second),
and $\Spix$
is the pixel sensitivity for a two--year mission. }
\label{tabfeat}
\begin{center}
\begin{tabular}{ccccccccc} 
\hline
\rule[-1ex]{0pt}{3.5ex}
  ${\nu}$  &
  channels &
  BW &
  FWHM &
  Orbit Time&
  Coverage&
  $\Npix$ &
  ${\sigma}_{1s}$ &
  $\Spix$ \\
  (GHz) &
   (\#) &
   ($^{\circ}$)&
   &
  (s) &
  (\%)&
  &
  (mK s$^{{1/2}}$) &
  ($\mu$K) \\
\hline
22 & 1 &  &  &  &   &  & 0.5  & $1.6$  \\
32 & 1 & 10\% & 7 & 5400 &  80 & 660 & 0.5  & $1.6$  \\
90 & 2 &  &  & &   &  & 0.57 & $1.8$  \\
\hline
\end{tabular}
\end{center}
\end{table}
%%%%%%%%%%%%%%%%%%%%%%%%%%%%%%%%%%%

\begin{figure*}
\centering
\includegraphics[width=0.3\hsize,angle=90]{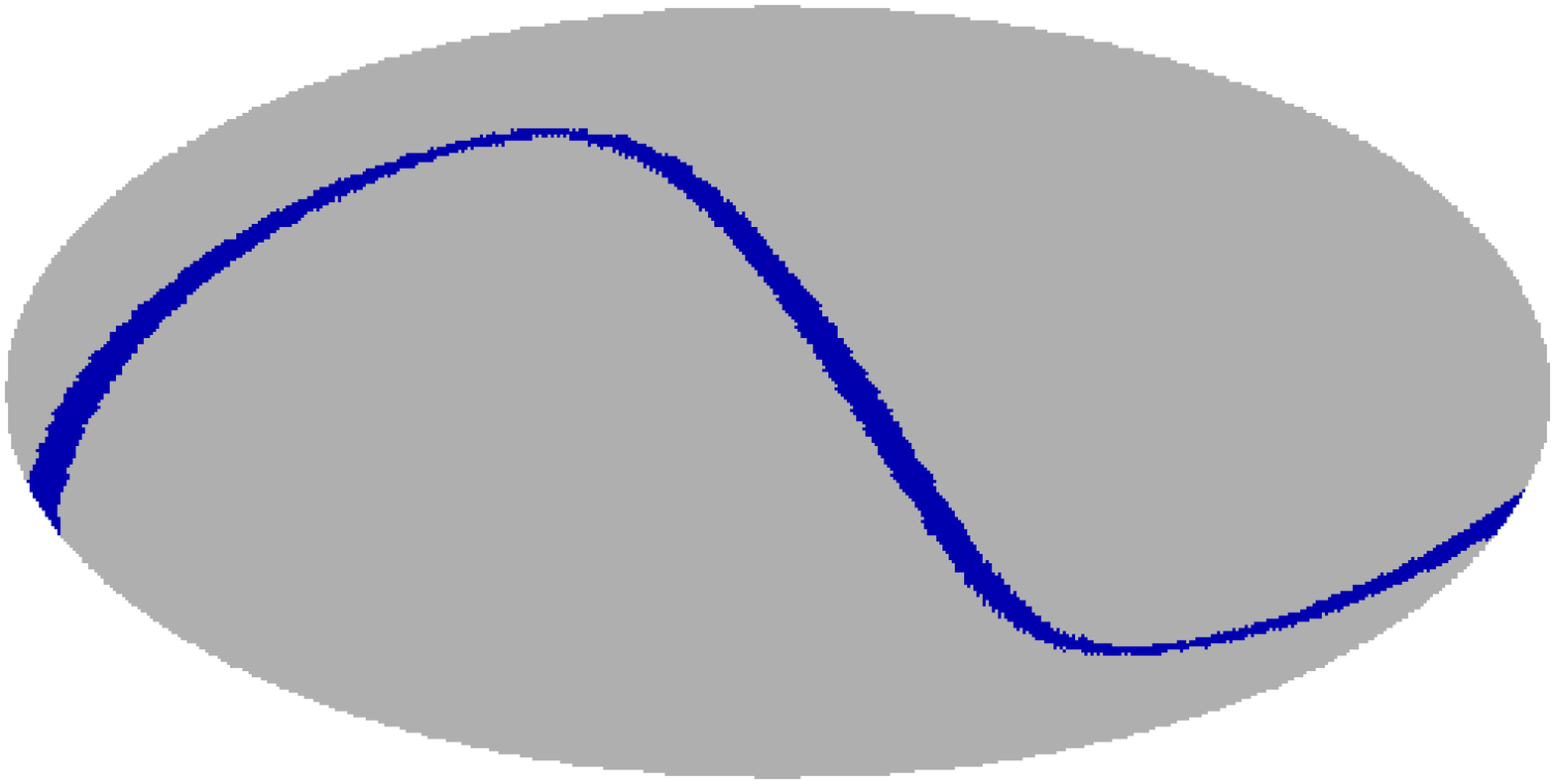}
\includegraphics[width=0.3\hsize,angle=90]{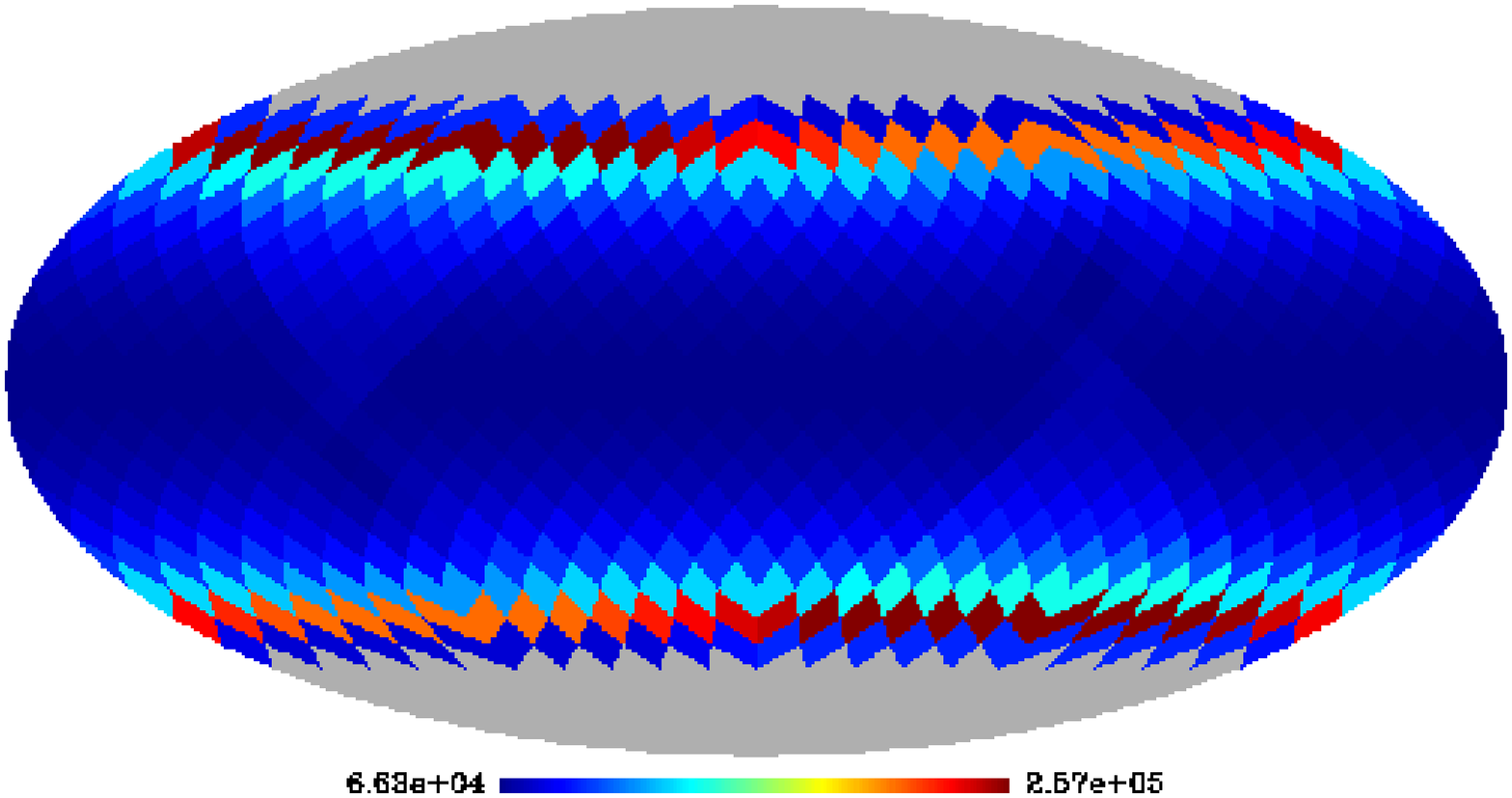}
\caption{Left: The sky, in Celestial coordinates, as scanned by SPOrt in
few orbits. 
Right: Pixel observing time (seconds) for two years of data 
taking. The pixel size is about $7^{\circ}$ 
(HEALPix parameter $Nside=8$).}
\label{orbitfig}
\end{figure*}

The SPOrt antennae, looking at the zenit, will cover 80\% of
 the sky thanks to the
motion of the Space Station,  whose orbit is tilted by 
$51.6^{\circ}$ with respect to the 
Celestial equator and is characterized by a period of about 5400~s
and a precession of about 70~days. This results in
the sky scanning pattern shown in Fig.~\ref{orbitfig},
providing a full coverage of the accessible region 
every 70 days. In Fig.~\ref{orbitfig} a map of the time spent over each 
pixel, of about 7$^{\circ}$ 
(HEALPix\footnote{http://www.eso.org/science/healpix} parameter
$Nside=8$), is also shown.

The main features of the SPOrt experiment, summarised in 
Table~\ref{tabfeat}, have 
been chosen to allow the accomplishment of SPOrt primary goals, 
e.g. a tentative 
detection of CMBR polarization on large angular scales, and the 
mapping of Galactic 
synchrotron emission. The former asks for both a nearly all--sky 
survey 
and a multifrequency approach to control  possible contaminations 
from Galactic 
foregrounds; the latter is expected from the channels at 22 and 
32~GHz.
 
The direct and simultaneous measurement of both  
Stokes parameters  $Q$ and $U$
optimizes the observing time efficiency, a remarkable improvement 
compared to other schemes providing either $Q$ or $U$ and thus reducing
it by a factor $2$. 

The SPOrt expected sensitivity to CMBR, resulting from the combination
of the four channels and taking into account the worsening due
to foreground subtraction (Dodelson 1997), is reported in Table~\ref{featCMB}.
In the frame of recent WMAP results on the optical depth of the Universe at 
the re--ionization epoch, SPOrt promises a solid detection of large scale CMBR
polarization.

%%%%%%%%%%%%%%%%%%%%%%
%%%%%%%%%%%%%%%%%%%%%%%%%%%%%%%%%%%%%%%%%%

%%%%%%%%%%%%%%%%%%%%%%%%%%%%%%%%%%%
\begin{table}
\caption{SPOrt sensitivity to CMBR, in thermodynamic temperature, 
combining the four channels and including Galactic synchrotron subtraction, 
for a two--year mission.}
\label{featCMB}
\begin{center}
\begin{tabular}{ccc} %% this creates nine columns
\hline
\rule[-1ex]{0pt}{3.5ex}
   $\sigma_{1s}$   &
   $\Spix$              &
   $\sigma$(P$_{\rm {rms}}$) \\
   (mKs$^{1/2}$)   &
   ($\mu$K)             &
   ($\mu$K) \\
\hline
 0.53  & $1.7$ & $0.15$ \\
\hline
\end{tabular}
\end{center}
\end{table}
%%%%%%%%%%%%%%%%%%%%%%%%%%%%%%%%%%%

\subsection {Instrument Design and Analysis}
\label{sec:instrdef}

The main concern in designing an experiment to measure
CMBR polarization is the low level of the expected
signal (1--10\% of the already tiny temperature anisotropies, 
depending on the scale), 
requiring specific instrumentation. 

The design optimization with respect to systematics generation, long term 
stability and observing time efficiency is indeed even more critical 
here than for experiments primarily designed to investigate CMBR 
temperature anisotropies. This is demonstrated by previous attempts to 
measure CMBR polarization where the final sensitivity was sizeably worse
than that expected from the istantaneous sensitivity of
the front--end noise (Keating \etal\ 2001; Hedman \etal\ 2002; 
Kogut \etal\ 2003). 
%%%%%%%%%%%%%%%%%%%%%%%%%%%%%%%%%
\begin{figure}
\centering
\includegraphics[width=0.8\hsize]{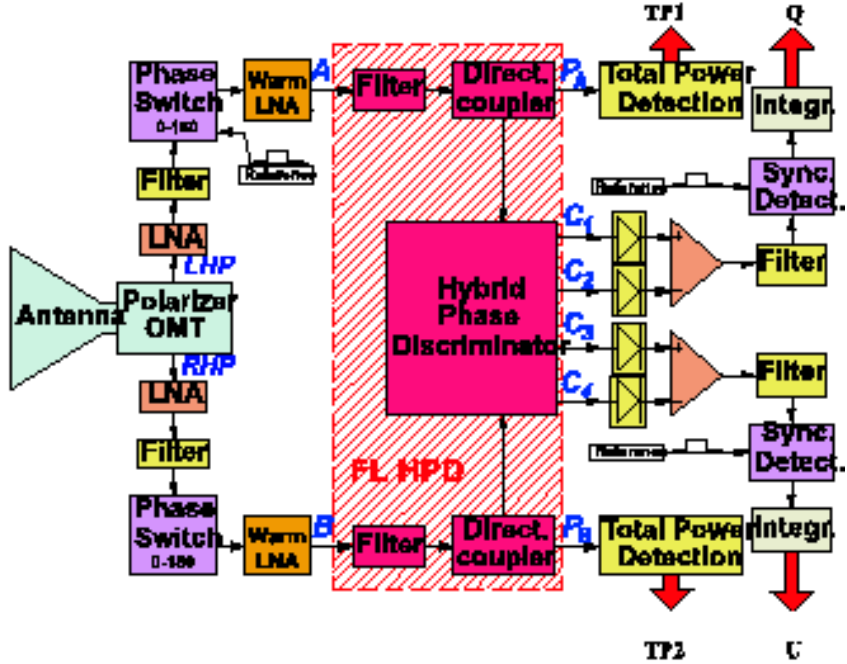}
\caption{Schematic block diagram of the SPOrt radiometers. Polarizer and OMT
extract the two circularly polarized components $LHP$ and $RHP$ collected by 
the horn. After amplification, the correlation unit (based on a Hybrid Phase 
Discriminator, see Peverini \etal\ 2002 for details) provides directly both 
$Q$ and $U$.}
\label{radioFig}
\end{figure}
%%%%%%%%%%%%%%%%%%%%%%%%%%%%%%%%%%

The following main choices were made 
while designing the polarimeter:
\begin{itemize}
\item correlation architecture to improve the stability;
\item correlation of the two circularly polarized components $E_L$ and 
$E_R$ to  directly and simultaneously measure both Q and U from the real and
imaginary parts of the correlation, respectively,
\begin{equation}
\begin{array}{rcr} 
Q \propto \Re (E_R E_L^*); &\,\,\,\,\,\,\,\,\,\, & 
U \propto \Im (E_R E_L^*)
\end{array} 
\end{equation} 
 while keeping 100\% efficiency;
\item detailed analysis of the correlation scheme to identify {\em critical}
      components and set their specifications in order to keep the 
      systematics at a level suitable to measure CMBR polarization;
\item custom development of components when state--of--the--art
      is not compliant with SPOrt requirements;
\item on axis simple optics (corrugated feed horns) to minimize the 
spurious polarization induced by both the $f$ pattern 
(see Carretti \etal\ 2001 for its definition) and the 
CMBR temperature anisotropy on the beam scale.
\end{itemize}
The equation giving the radiometer sensitivity 
(Wollack 1995; Wollack \etal\ 1998) 
\begin{equation}
  \Delta T_{\rm rms} = \sqrt{
                      {k^2 \,T_{\rm sys}^2 \over \Delta\nu\,t}+
               T_{\rm offset}^2
               \left({\Delta G \over G}\right)^2 +
               \Delta T_{\rm offset}^2
            }
\label{trmseq}
\end{equation}
can help us find the parameters to be
controlled to minimize the systematics. 
Here $T_{\rm sys}$, $T_{\rm offset}$ and $\Delta
T_{\rm offset}$ represent the system
temperature, the offset equivalent temperature and its fluctuation,
respectively; $G$ is the radiometer
gain, $t$ the integration time, $\Delta \nu$ the radiofrequency bandwidth
and $k\simeq 1$ a constant depending on the radiometer type.

The first term in Eq.~(\ref{trmseq}) represents the white noise of an 
ideal and stable radiometer. Its contribution can be reduced by cooling 
to cryogenic temperature the front--end (first stage of the Low 
Noise Amplifiers (LNAs), polarizer and the OMT).
Unfortunately, passive cooling is not feasible
for low orbits because of variations of the Sun illumination. 
For SPOrt,  a mechanical pulse--tube cryocooler was adopted, 
ensuring high cooling efficiency  down to $\sim$80--90~K. The 
needed temperature stability ($\pm 0.1$~K) is
achieved by a closed--loop active control.
An active control is also adopted for the warm parts (Horn and back--end),
allowing a temperature stability of $\pm0.2$~K.

The second and the third term of Eq.~(\ref{trmseq}) represent 
the contributions of gain and offset fluctuations, respectively,
 generated by instrument instabilities.
It is clear that the noise behaviour can 
be kept close to the ideal (white) case 
provided the offset is kept under due control.\\

A scheme of the SPOrt radiometers is sketched in  
Fig.~\ref{radioFig}.
The Polarizer and the
OMT extract the two circularly polarized components $E_L$~and~$E_R$
collected by the dual--polarization feed horn. 
After amplification, the two components are
correlated by the Correlation Unit (CU), consisting of 
an Hybrid Phase Discriminator (HPD), four diodes and two differential 
amplifiers. In details, the HPD generates the four outputs:
\begin{eqnarray}
C_1 &\propto& E_L + E_R \\
C_2 &\propto& E_L - E_R \\
C_3 &\propto& E_L + j\,E_R \\
C_4 &\propto& E_L - j\,E_R 
\end{eqnarray}
which are square--law detected by the four diodes
\begin{eqnarray}
\left|C_1\right|^2 &\propto& \left|E_L\right|^2 + \left|E_R\right|^2 + 2\Re (E_R E_L^*) \\
\left|C_2\right|^2 &\propto& \left|E_L\right|^2 - \left|E_R\right|^2 - 2\Re (E_R E_L^*) \\
\left|C_3\right|^2 &\propto& \left|E_L\right|^2 + \left|E_R\right|^2 + 2\Im (E_R E_L^*) \\
\left|C_4\right|^2 &\propto& \left|E_L\right|^2 - \left|E_R\right|^2 - 2\Im (E_R E_L^*). 
\end{eqnarray}
The differences performed by the two differential amplifiers provide
\begin{equation}
\begin{array}{lllll}
\left|C_1\right|^2 -\left|C_2\right|^2 &\propto& \Re(E_R E_L^*) &\propto& Q \\
\left|C_3\right|^2 -\left|C_4\right|^2 &\propto& \Im(E_R E_L^*) &\propto& U
\end{array}
\end{equation}
allowing the simultaneous measurement of the two Stokes parameters 
$Q$~\&~$U$.
A nice feature is that no effort is needed to  equalize the mean phase
difference between $E_R$ and $E_L$ when propagating the signal through 
the instrument,
since an error on this difference simply results in a 
polarization angle rotation, well
recoverable in the calibration procedure. 
Indeed, linear polarization might also be measured by correlating
the two linear polarizations $E_x$,~$E_y$. However, this would only provide 
one linear Stokes parameter ($U$), reducing the experiment efficiency,
and would generate  a
combination of $U$ and $V$ in the output in case of errors in the 
phase difference. Therefore, great care would be needed in equalizing 
the path of the two components. 

Our analysis to track down critical components for the 
offset generation pointed out important sources in both the CU
and the antenna system (horn, polarizer and OMT).

We found that the CU needs an HPD with high rejection of the unpolarized
component, at a level well beyond available instrumentation.  A custom
device (Tascone \etal\ 2002; Peverini \etal\ 2002)
providing $> 30$~dB rejection was thus developed. In combination with a 
lock--in system, this makes the CU contribution to the offset negligible,
the total rejection being $\simeq 60$~dB.

The contribution to the offset coming from the antenna system 
is given by (Carretti~\etal\ 2001):
\begin{equation}
 T_{\rm offset}   =  S\!P_{\rm OMT}\left(T_{\rm sky} +
                 T_{\rm noise}^{\rm Ant}
               \right)
                  +
                  S\!P_{\rm pol}
                  \left(T_{\rm sky} +
                            T_{\rm noise}^{\rm horn} -
                            {T_{\rm ph}^{\rm pol} \over
                     \eta}
                          \right),
                  \label{AB0TNeq}
\end{equation}
where $T_{\rm sky}$ is the signal coming from the sky,
$T_{\rm noise}^{\rm horn}$ is the noise generated by the horn alone,
  $T_{\rm noise}^{\rm Ant}$
is the noise temperature of the whole antenna system, $\eta$ is the
efficiency of the feed horn and $T_{\rm ph}^{\rm pol}$ is the physical
temperature of the polarizer. The two quantities
\begin{eqnarray}
 S\!P_{\rm OMT} & = & 2\,{\Re(S_{A1}S_{B1}^*)\over \left|S_{A1}\right|^2},
                     \\
 S\!P_{\rm pol} & = & {1\over 2} \left(1 - {\left|S_{\perp}\right|^2\over
                                 \left|S_{\parallel}\right|^2}\right),
                                 \label{sppoleq}
\end{eqnarray}
describe the performances of the OMT and 
the polarizer, respectively, in terms of offset generation.
Uncorrelated signals like noise and sky emission are partially
detected as correlated because of the OMT
cross--talk ($S_{A1}$ and $S_{B1}$ representing the transmission and cross--talk
coefficients of the OMT, respectively) and the polarizer
attenuation difference ($S_{\parallel}$ and $S_{\perp}$ giving the 
transmissions of the two polarization states).

To set specification requirements for our OMT and polarizer
we considered the radiometer knee frequency, $f_{\rm knee}$. This
parameter provides the time
scale at which the low frequency  component of the
noise, also known as 1/$f$, prevails on the white one, and is therefore
suitable to quantify radiometer instabilities.

%%%%%%%%%%%%%%%%%%%%%%%%%%%%%%%%%
\begin{figure} 
\centering
\includegraphics[width=0.9\hsize]{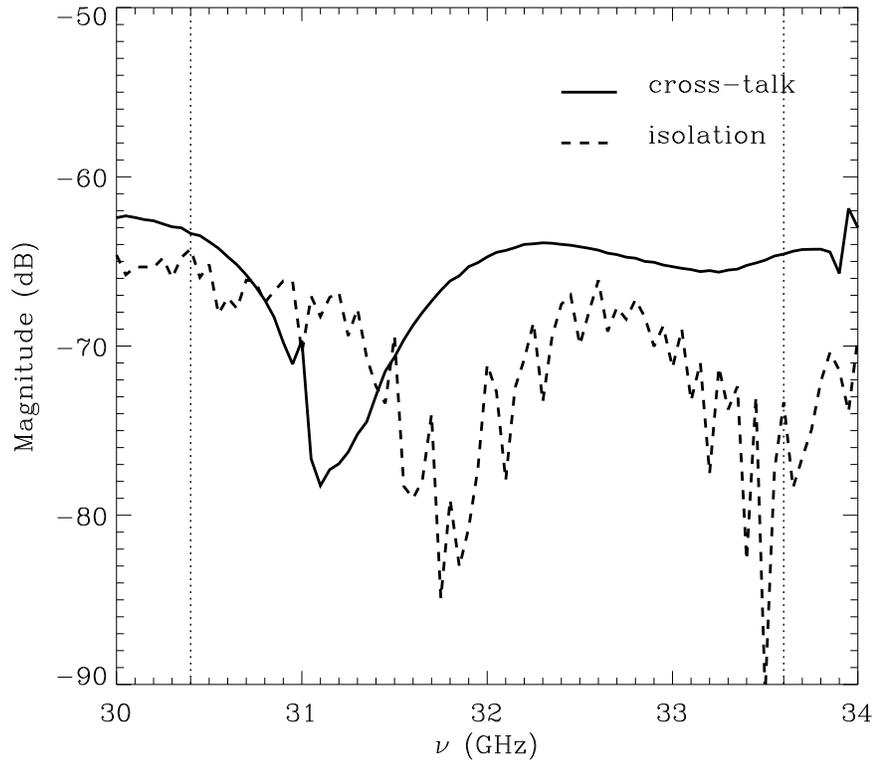}
\caption{Magnitude of both the isolation between the two rectangular ports
and the cross--talk between
the two polarization states for the 32~GHz OMT of SPOrt. The vertical dotted lines
define the 10\% band.}
\label{crossfig}
\end{figure}
%%%%%%%%%%%%%%%%%%%%%%%%%%%%%%%%%%

As is known, destriping techniques can remove most of 
the effects of  1/$f$ noise
provided the radiometer knee frequency is lower than
the signal modulation frequency (see Sect.~\ref{sec:destr})
that, for SPOrt, corresponds to
the orbit frequency $f_{\rm o} = 0.18$~mHz.

Currently available InP LNA have rather high
knee frequencies  ($f_{\rm knee}^{\rm lna}\sim$ 100--1000~Hz).
However,  the knee frequency of a correlation receiver is related to that
of its amplifiers by the formula:
\begin{equation}
f_{\rm knee} = \left({T_{\rm offset}\over T_{\rm sys}}\right)^{2/\alpha}
                 f_{\rm knee}^{\rm lna}
\end{equation}
where $T_{\rm offset}$ is the radiometric offset, $T_{\rm sys}$ the system
temperature and $\alpha\sim 1$ (Wollack 1995) is the power--law 
index of the LNA's $1/f$ noise behaviour.
%%%%%%%%%%%%%%%%%%%%%%%%%%%%%%%%%
\begin{figure}
\centering
\includegraphics[width=1.00\hsize]{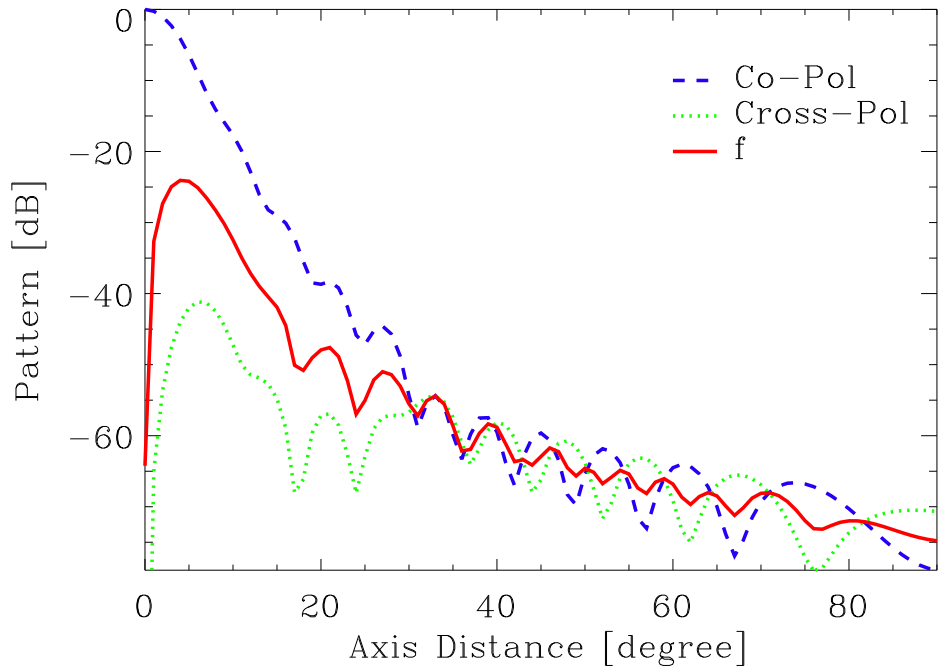}
\caption{Co--polar, cross--polar and $f$ patterns, with respect to
the axis distance $\theta$, for the 90~GHz feed horns of SPOrt . The spurious
polarization coming from the horn is generated by the anisotropy distribution 
$\Delta T(\theta\phi)$ in combination with the $f$ pattern (see text).}
\label{feedHornFig}
\end{figure}
%%%%%%%%%%%%%%%%%%%%%%%%%%%%%%%%%%

To match the condition for a successful destriping
($f_{\rm knee} < f_{\rm o}$, with $f_{\rm knee} = 0.1 f_{\rm o} = 0.018$~mHz
 as a goal),
SPOrt requirements on the OMT cross--talk and the difference between the 
attenuations of the two polarization states were quantified in $-60$~dB 
and $-30$~dB, respectively. This guarantees an offset value 
as low as $T_{\rm offset}\sim$~50~mK  which, combined with a
$T_{\rm sys} \sim$~100~K, translates into  the knee frequency
\begin{equation}
f_{\rm knee} \sim 2.5\times 10^{-7} f_{\rm knee}^{\rm lna}
\end{equation}
fully satisfying the requirements for an efficient destriping.

Commercially available state--of--the--art OMTs have isolation
values not better than $-40$~dB, which forced us to develop new
 hardware. Fig.~\ref{crossfig} shows the extremely good results
already obtained 
for the 32~GHz channel: the cross--talk is as low as $-65$~dB, with an
 isolation of about 70~dB.

Besides the offset generation, a further worrysome source of systematic errors
is the spurious polarization generated by the optics (Carretti \etal\ 2001).
This is due to the anisotropy distribution of the unpolarized radiation,
modulated by the $f$ pattern:
\begin{eqnarray}
\Delta T^{\rm horn} &=&{1\over\Omega_A}\int_0^{\pi}\sin\theta\,d\theta
                                    \int_0^{\pi/2}
                                   \left[\Delta T_b(\theta,\phi)\right.\nonumber\\
             &-&                         \Delta T_b(\theta,\phi+\pi/2)+
                                         \Delta T_b(\theta,\phi+\pi)\nonumber\\
             &-&                   \left.\Delta T_b(\theta,\phi+3/2\pi)
                                           \right]\, f(\theta,\phi)\,d\phi\,,\\
             & &\nonumber\\
f(\theta,\phi) &=& -P(\theta, \phi)\chi^*(\theta, \phi+\pi/2) \nonumber\\
               &+& \chi(\theta, \phi) P^*(\theta, \phi+\pi/2)\,,
\end{eqnarray}
where $P$ and $\chi$ are the co--polar and cross--polar patterns, 
normalized to the P maximum, respectively,
and $\Omega_A$ is the antenna beam.
The $f$ pattern consists of a 4--lobe structure, with lobe size close to
the FWHM of the instrument. 
A worst--case analysis of the contamination gives
\begin{equation}
\Delta T^{\rm horn} = S\!P_{\rm horn}\,\, \Delta T_{\rm rms} ({\rm FWHM})\, ,
\end{equation}
with
\begin{equation}
 S\!P_{\rm horn} = 2\,{1\over\Omega_A}\int_0^{\pi}\sin\theta\,d\theta
                                    \int_0^{\pi/2}\,d\phi\,f(\theta,\phi)
\end{equation}
and $\Delta T_{\rm rms} ({\rm FWHM})$ the rms temperature anisotropy on FWHM
scale.
As shown in Fig.~\ref{feedHornFig},
in case of the SPOrt feed horns 
the contribution of the $f$ pattern is
$S\!P_{\rm horn}\sim -24$~dB and the {rms} contamination due to the 
30~$\mu$K--CMBR anisotropy  is thus lower than 0.2~$\mu$K.

\subsection{Calibration procedure}

The accuracy needed for measuring CMBR polarization requires good methods for
calibrating the response of the instrument to small polarized signals.
Furthermore, in the absence of  well--characterized astrophysical sources, 
specialized techniques are needed to inject calibration markers. 
In fact, standard marker injectors are not suitable for calibrating 
tensorial quantities as the ($Q$,~$U$) pair measured by SPOrt.
A new concept calibrator,
valid for any radio--polarimeter and based on the insertion
of three signals at different position angles, has thus been developed
(Baralis \etal\ 2002).

To focus the problem it is convenient to represent the entire radiometer
in terms of a system with two input and output signals. The two
inputs $A$ and $B$ are high frequency signals in the circular
polarization base ($E_L$ and $E_R$), whereas the two outputs $Q_m$
and $U_m$ are the low frequency signals corresponding to the
Stokes parameters directly measured by the radiometer. According
to this model, the output of the radiometer can be described by
the following matrix expression:
 \beq
 \qmatrix{Q_m\cr U_m}=\qmatrix{H_{qq} & H_{qu}\cr H_{uq} &
 H_{uu}}\qmatrix{Q\cr U}+\qmatrix{T_{qa} & T_{qb} \cr T_{ua} &
 T_{ub}}\qmatrix{|A|^2 \cr |B|^2}
 \label{eq01}
 \eeq
where $Q$ and $U$ are the input Stokes parameters.
The matrices $\unun{H}$ and $\unun{T}$
in Eq.~(\ref{eq01}) are generic, but real, since they deal with
quadratic quantities. Moreover, they transform polarization circles in the
$QU$--plane into rotated
and translated ellipses. Under the reasonable assumption that the
intensities of the total signals $A$ and $B$ are equal
($|A|^2=|B|^2\propto P$), Eq.~(\ref{eq01}) can be rewritten
as $\un{Y}=\unun{H}\,\un{X}+\un{C}\,P$, where $\un{C}$ is a
two--element column vector. The calibration procedure, i.e. the
evaluation of the matrices $\unun{H}$ and the vector $\un{C}$,
can be
accomplished by directly measuring the quantities $Q_m$ and $U_m$
in presence of predefined signals (markers). In fact, by injecting
at different times three markers with known polarization
angles $\theta_i$, each corresponding to a linearly
polarized field, and by
detecting the corresponding output variations, one ends up with a
matrix equation to be solved for the matrix $\unun{H}$:
 \begin{equation}
 \unun{H}\qmatrix{\frac{\Delta\un{X}_3}{\Delta P_3}-\frac{\Delta\un{X}_1}{\Delta P_1}
 \frac{\Delta\un{X}_2}{\Delta P_2}-\frac{\Delta\un{X}_1}{\Delta
 P_1}}= 
\qmatrix{\frac{\Delta\un{Y}_3}{\Delta P_3}-\frac{\Delta\un{Y}_1}{\Delta P_1}\;, &
 \frac{\Delta\un{Y}_2}{\Delta P_2}-\frac{\Delta\un{Y}_1}{\Delta
 P_1}}
 \label{eq03}
\end{equation}
where $\Delta\un{X}_i$, $\Delta\un{Y}_i$ and $\Delta P_i$ are the
variations of the input and output Stokes parameters and of the
total power, respectively. To obtain a well--conditioned matrix the markers
should correspond to signals with a relative rotation of either 45$^{\circ}$
or 60$^{\circ}$.

\subsubsection{Marker injector}

The calibration procedure described in the previous section
requires the injection of different markers in the radiometer,
just behind the antenna. Unfortunately, this procedure must be
carried out in operative conditions, e.g. in presence of both 
polarized and unpolarized radiation. Therefore, the marker
injector must degrade the relevant signals as little as possible.
In particular, this device has to exhibit a high return loss and
prevent depolarization of the incoming signals.
\begin{figure}
\centering
\includegraphics[width=0.5\hsize]{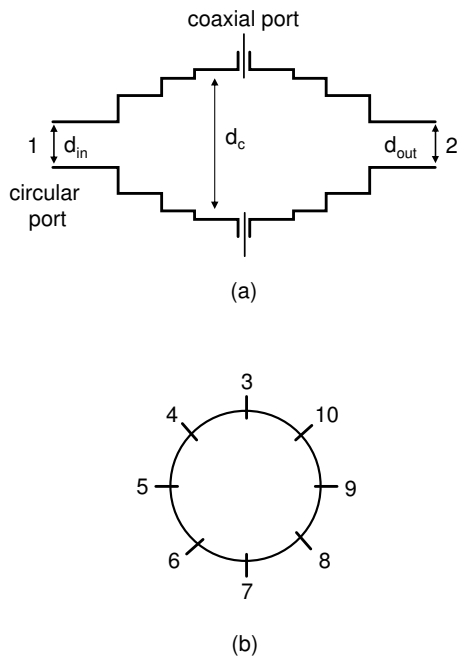}
\caption{Scheme of the marker injector. a) Longitudinal cut. 
b) Cross--section of the central block.}
\label{fig01}
\end{figure}
\begin{figure}
\centering
\includegraphics[width=0.9\hsize]{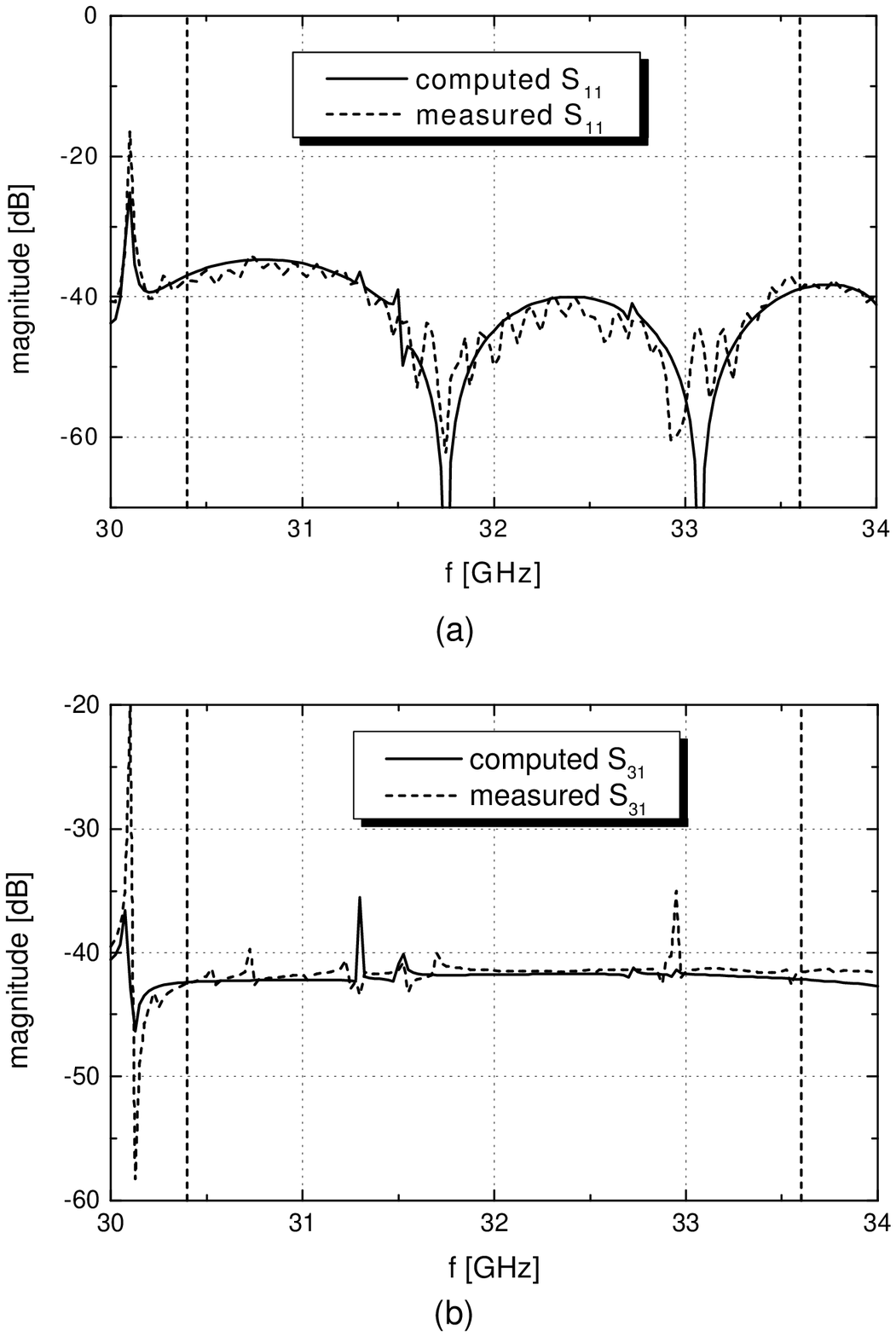}
\caption{Measured an computed scattering parameters of the Ka--band marker 
injector. Top: Reflection coefficient. Bottom: Coaxial/circular coupling 
factor. Vertical dashed lines define the SPOrt band.}
\label{fig02}
\end{figure}

The design of the Ka--band prototype is depicted in
Fig.~\ref{fig01}, 
and consists of three blocks (Peverini \etal\ 2002b).
The central block is formed by an enlarged circular waveguide in which
eight K--connectors are inserted. Although only three markers have
to be injected inside the circular waveguide, all the eight
connectors are necessary to preserve the azimuthal symmetry of the
structure and, hence, to avoid depolarization of the incoming
signals. The coupling level between coaxial and circular ports can
be controlled by adjusting the penetration length of the internal
wire of the K--connectors. The two lateral blocks provide a
matching structure between the input/ouput circular waveguides and
the enlarged central waveguide. The matching section is formed by
a cascade of several circular steps. The device was analyzed by
the Generalized Scattering Matrix approach (Itoh 1989) 
and designed by
evolution--strategy methods. All the components were manufactured by
highly precise electrical--discharge techniques.

In Fig.~\ref{fig02} 
the measured and simulated reflection
coefficient $S_{11}$ at the circular port for the fundamental
TE$_{11}$ mode and the coupling $S_{31}$ between circular
and coaxial ports are reported. The measured coaxial/circular
coupling factor is $\approx-42\,\rm dB$ in all the band of
interest ($30.4 \div 33.6\,\rm GHz$) and the fundamental mode 
at the circular port
exhibits a reflection coefficient better than $- 35\,\rm dB$ in
the same band. The measured in--band insertion loss 
 is about 0.02~dB, which can be correctly
predicted by using an equivalent surface resistance of
$10\,\rm\mu\Omega cm$.

\subsection{Destriping and Map Making}
\label{sec:destr}

Low frequency noise 
induces correlations among successive samples of the measured signal
and leads  to typical striping effects when producing sky maps.

However, when data are taken from  spinning spacecrafts, most 
of the low frequency noise can be removed by software, provided the
radiometer knee frequency  is lower than  the satellite spin frequency
(Janssen \etal\ 1996). As detailed in Sect.~\ref{defaps},
the low frequency noise of the SPOrt radiometers
is expected to have a $1/f$--like spectrum  
dominated
by transistor gain fluctuations. The present state of the
instrument already guarantees a knee frequency
$f_{\rm knee}$ lower than the ISS orbit frequency $f_o$.

Various destriping algorithms have been proposed in recent
years to clean CMBR anisotropy data (e.g. Delabrouille 1998; Maino \etal\ 1999), 
together with a first
algorithm specifically designed for the polarization case (Revenu \etal\ 2000).
They are generally based on $\chi^2$ minimizations and involve
large matrix inversions.

A different technique has been implemented  for SPOrt (Sbarra \etal\ 2003),
consisting of a simple but effective iterative algorithm relying
upon minimal assumptions: the radiometer must
be stable during the signal modulation period (the time needed to complete 
one orbit in case of SPOrt), so that the
instrumental offset can be assumed to be constant over the same period,
and there must be enough overlap between different orbits.
In such a situation the noise
can be split into two parts: for time scales shorter than the orbit
period it is essentially white, whereas for longer timescales
the $1/f$ component can be modelled as a different
constant offset for each orbit. A simple iterative procedure can  then be
applied to remove these offsets from the Time Ordered Data (TOD) before
map--making. No assumptions need to be made about the statistical properties
of the noise. The map--making itself consists of a simple average of the
measurements corresponding to the same pixel, which is the optimal
method once only white noise is left (Tegmark 1997).

Although the algorithm has been studied to destripe  $Q$ and $U$
data, it can be easily simplified to deal with
scalar quantities. The average of the measured signal is lost 
in the latter case.
However, for maps of polarization data like $Q$ and $U$,
the average signal can still be kept provided the polarimeter
reference frame rotates about the standard reference frame (polar basis,
see Berkhuijsen 1975) 
while running along each orbit, as in case of SPOrt. 
This is a nice feature, especially
when measuring foreground contributions.

\begin{figure}
\centering
\includegraphics[width=0.3\hsize,angle=90]{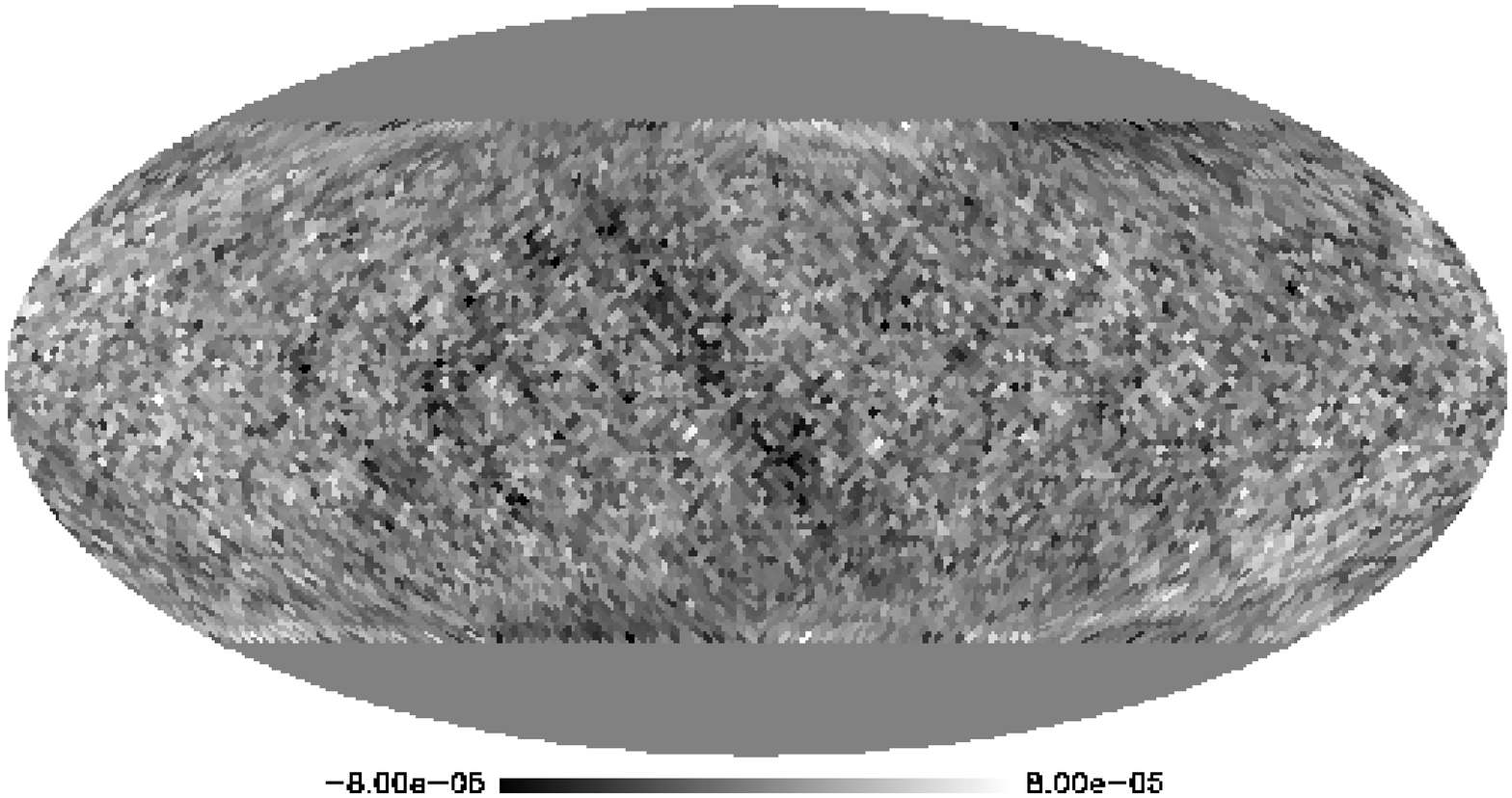}
\includegraphics[width=0.3\hsize,angle=90]{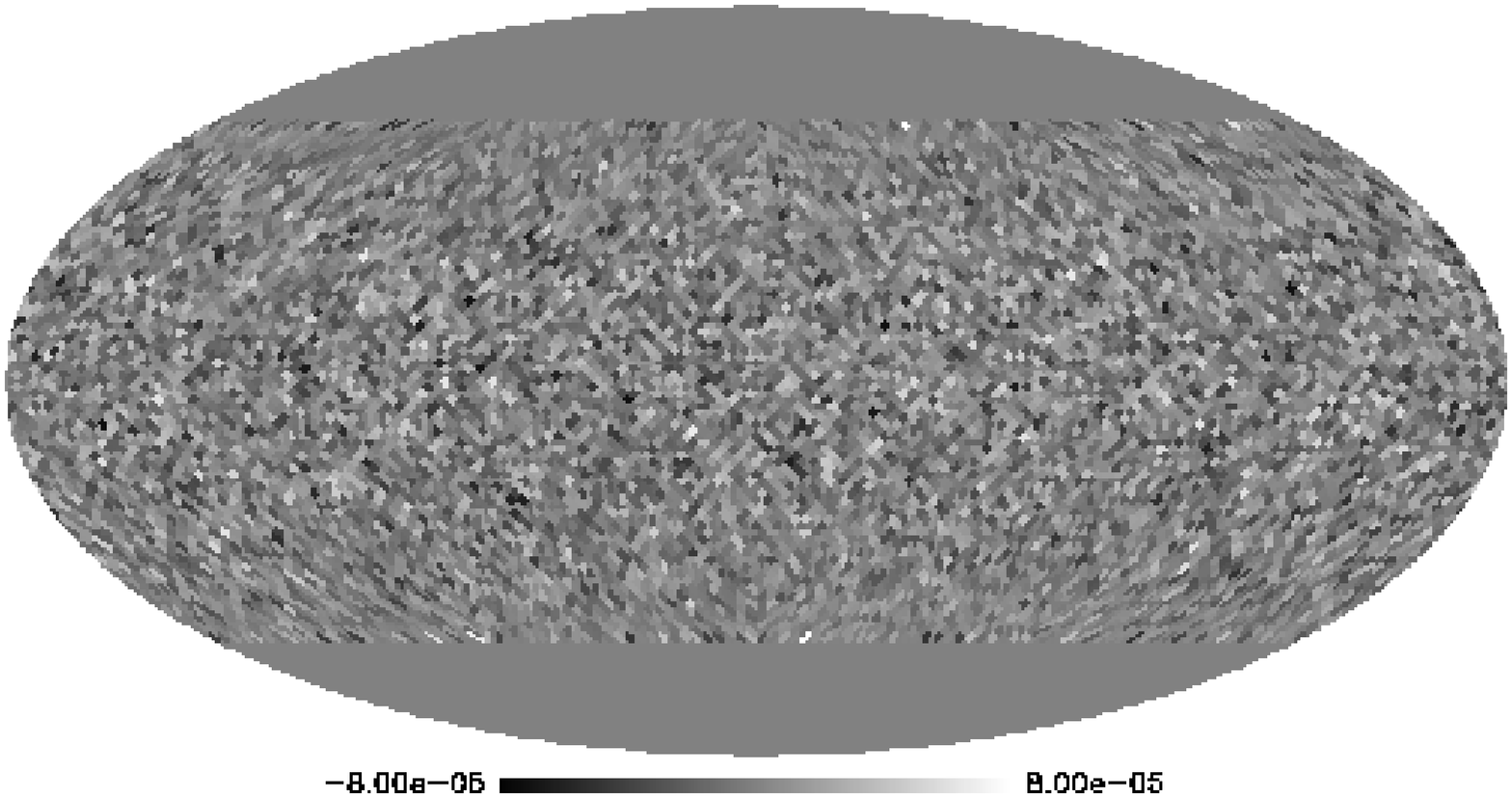}
\caption{Noise simulated maps before (top) and after (bottom) destriping, for
the case $f_{\rm knee}=1.8\cdot 10^{4}$, in Kelvin. The 
HEALPix parameter $Nside$ is 32 (courtesy of A\&A).}
\label{mapdes}
\end{figure}

The good performances of the technique are made evident
in Fig.~\ref{mapdes} where we show a simulated noise
map (both $1/f$ and white) before and after destriping.
%%%%%%%%%%%%%%%%%%%%%%
%%%%%%%%%%%%%%%%%%%%%%%%%%%%%%%%%%%%%%%%%%
%%%%%%%%%%%%%%%%%%%%%%%%%%%%%
\begin{table}[htb]
\caption{
Excess rms noise due to low--frequency contributions,
with respect to the white noise level, for pixels of $\simeq 7^{\circ}$.}
\begin{center}
\begin{tabular}{lll}
\hline
$f_{\rm knee}$~(Hz) & Before Destriping &  After Destriping \\
\hline
$1.8\times 10^{-4}$ & 310\% & 6\% \\
$1.8\times 10^{-5}$ & 35\% & $<1$\% \\ \hline
\end{tabular}
\end{center}
\label{rmsnoise}
\end{table}
%%%%%%%%%%%%%%%%%%%%%%
%%%%%%%%%%%%%%%%%%%%%%
\begin{figure}
\centering
\includegraphics[width=1.1\hsize]{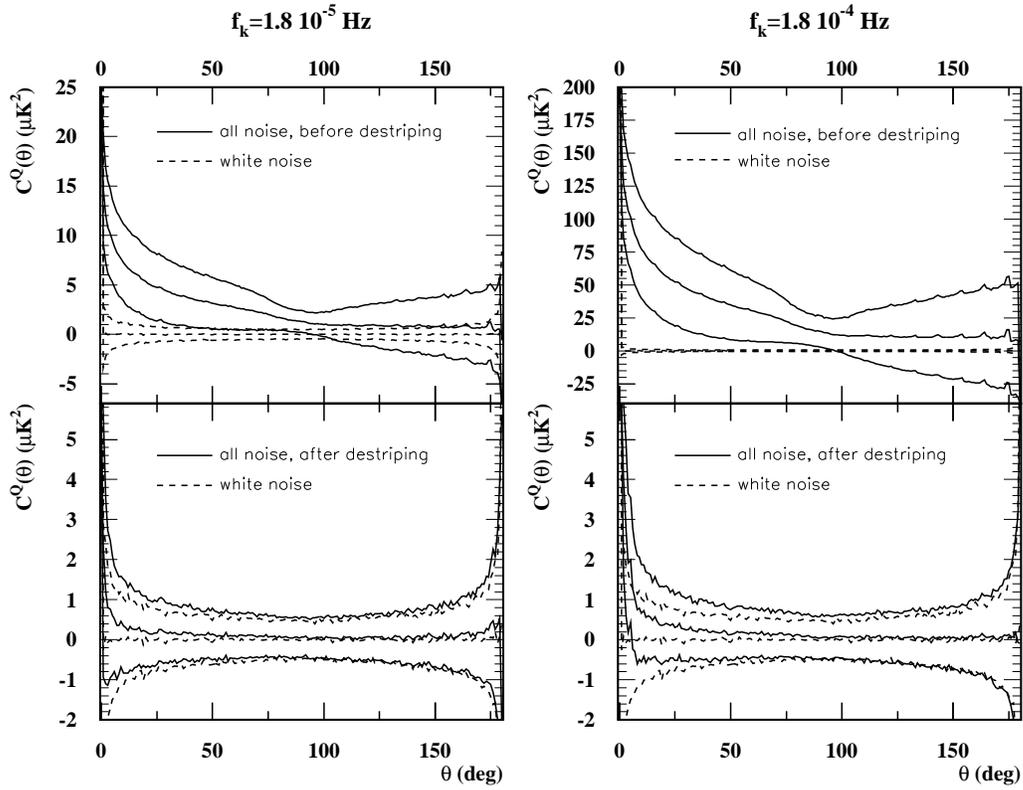}
\caption{Average and 1$\sigma$ band of  correlation functions $C^Q(\theta)$
 measured from 500 noise maps,
before and after destriping, for two
values of the knee frequency $f_{\rm knee}$.
Before destriping the $y$ scale is not the same for the
two $f_{\rm knee}$ values.
Each simulation
corresponds to about one year of SPOrt realistic data taking.
The case of purely white noise is shown for comparison (courtesy of A\&A).}
\label{fig:noisecorr}
\end{figure}
%%%%%%%%%%%%%%%%%%%%%%%%%%%%%%
%%%%%%%%%%%%%%%%%%%%%%%%%%%%%%%%%%%%%%%%%%%%%%
\begin{figure}
\centering
\includegraphics[width=\hsize]{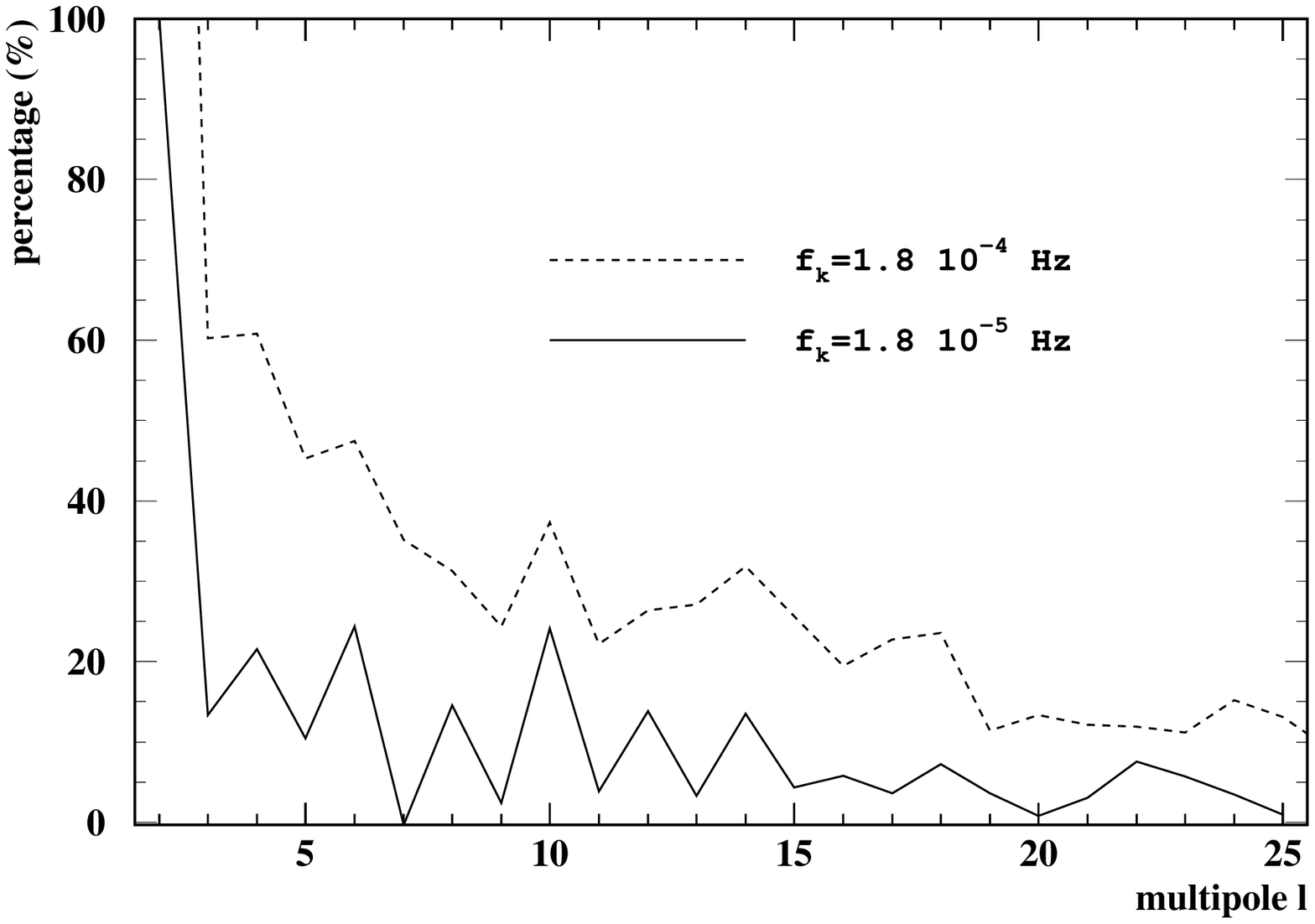}
\caption{Increment, due to the presence
of residual correlated noise, in the  rms of measured
noise power spectra, after destriping, as a function of
the multipole, in percentage of the rms of purely
white--noise power spectra (courtesy of A\&A). }
\label{fig:clserror}
\end{figure}
%%%%%%%%%%%%%%%%%%%%%%%%%%%%%%%%%%%%%%%%%%%%%%
One possible way to quantify the quality of the destriping is measuring
the fractional excess pixel noise with respect to the
case of purely white noise. Results are shown in Table~\ref{rmsnoise}
for two different values of
the knee frequency, corresponding to the SPOrt goal knee frequency,
$f_{\rm knee}=1.8\times 10^{-5}$~Hz, and the SPOrt orbit frequency,
$f_{\rm o}=1.8\times 10^{-4}$~Hz,
the latter representing a conservative case.

Another way to quantify the residual correlated noise is
measuring and inspecting the two--point correlation
functions C$^{Q}(\theta)=\langle Q(1)Q(2)\rangle$ and 
C$^{U}(\theta)=\langle U(1)U(2)\rangle$ (see Sect.~\ref{defaps}) of
simulated $Q$ and $U$ noise maps.
Averages and 1$\sigma$ bands of 500 correlation functions
C$^{Q}(\theta)$ 
measured from maps containing both white and $1/f$ noise, before and
after destriping, are compared
to the purely white noise case in Fig.~\ref{fig:noisecorr} for the
same knee frequencies as in the previous test.
As expected, the correlated noise is strongly reduced by the destriping
procedure, the residuals falling within the statistical error of
the white noise case for  $f_{\rm knee}=f_{\rm knee}^{\rm goal}$.

The polarization power spectra $C_{El}$ and $C_{Bl}$ can be obtained from 
the measured correlation functions
by inverting Eqs.~(\ref{t1u2}) (Kamionkowski \etal\ 1997b; Zaldarriaga 1998;
Ng \& Liu 1999). 
If the correlation functions are measured directly on $Q$ and $U$  maps
via $O(\Npix^2)$ operations ($\Npix$ being the number of
pixels in the measured map), this method (Sbarra \etal\ 2003) has the 
advantage of avoiding  edge problems (Chon \etal\ 2003) arising when 
using fast spherical transforms. 

The region of low multipoles is the most sensitive to
low frequency residuals, some contributions being always found here
even after the application of other destriping techniques (Maino \etal\ 1999).
Even though residual noise correlations can be modelled and 
subtracted from the measured 
$C^{Q,U}(\theta)$  functions before integration, their presence translates 
into an increment of the rms of the measured power spectra. A rough 
estimate for the case
of SPOrt is shown in Fig.~\ref{fig:clserror}.

\section{Astrophysics and Cosmology with SPOrt}
\label{phys}

%%%%%%%%%%%%%%%%%%%%%
\begin{figure}
\label{gianni}
\centering
\includegraphics[width=0.5\hsize,angle=90]{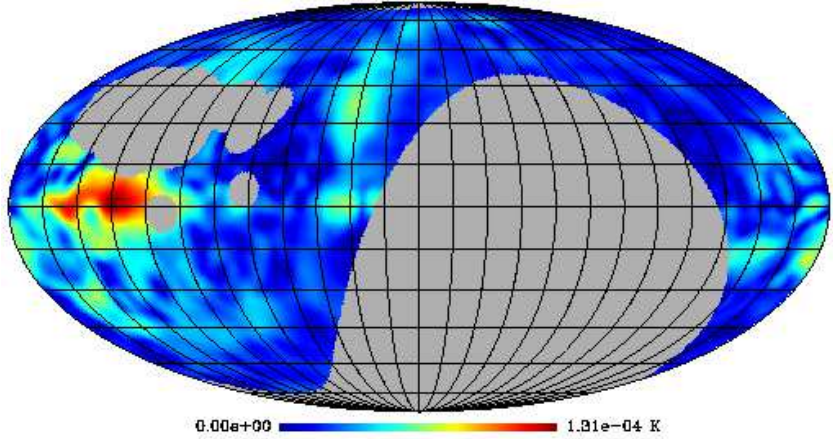}
\caption{Synchrotron template at 22 GHz (see Sect.~\ref{templat}), 
featuring a $P_{\rm rms}=17 \mu$K.}
\end{figure}
%%%%%%%%%%%%%%%%%%%%%%%%%%%%%%%%%%%%%%%%%%

A first goal that SPOrt will achieve is a deeper knowledge of the
polarization of the Galactic signal, in the microwave band. In
particular, according to  the synchrotron template 
described in Sect.~\ref{templat}, which we believe to be
not too far away from reality, SPOrt should be able to produce polarized 
Galaxy maps at both 22 and 30~GHz. 
In fact, as shown in Fig.~17, the synchrotron polarized emission,
as seen by the $7^{\circ}$--FWHM SPOrt optics at 22~GHz, is expected to
have a peak of about $100\,~\mu$K, two orders of magnitude 
larger than the experimental pixel sensitivity
in the same channel ($\Spix=1.6~\mu$K);
furthermore, the average of the polarized signal is $P_{\rm rms}=17~\mu$K.
Assuming a power--low dependence on the frequency with index
$\beta_{\rm synch}=-3$, the polarized intensity 
at 30~GHz is then expected to be just a factor 2--3 lower than at 22~GHz, 
so that SPOrt is likely to provide a Galaxy map at this frequency as well.

\begin{figure}
\centering
\includegraphics[width=0.5\hsize,angle=90]{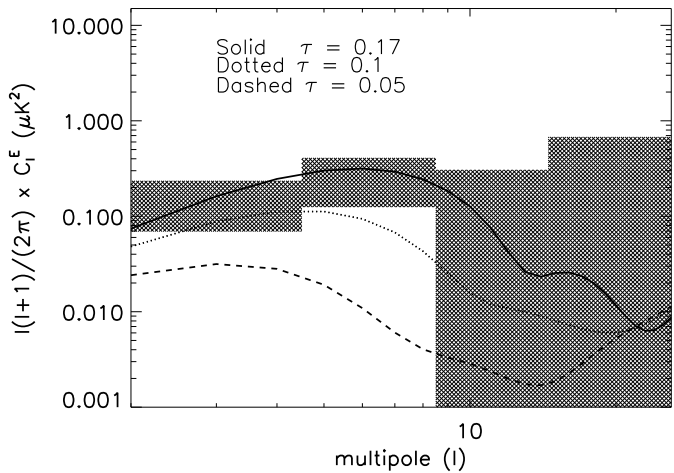}
\caption{$1-\sigma$ error bands, as expected from SPOrt measurement of 
the $C_{El}$ power spectrum,
including cosmic variance, considering the sensitivity
quoted in Table~\ref{featCMB}
and $\tau=0.17$, as recently found by WMAP. APS corresponding
to $\tau=0.1$ and $\tau=0.05$ are shown for comparison.}
\label{SportSens}
\end{figure}

The SPOrt pixel sensitivity quoted in Table~\ref{featCMB} does not 
allow the building of CMBR polarization maps.
However, a CMBR band power spectrum as that shown 
in Fig.~\ref{SportSens} is expected to be provided 
if the recent WMAP results are confirmed. 1--$\sigma$ error bands
are calculated following the recipe provided by Tegmark (1997).
The plot also suggests that a non--null (at 95\%C.L.) measurement of 
the mean polarized signal $P_{rms}$  
might be provided, already  for $\tau\simeq 0.1$, by 
full--sky statistical analyses
like the Maximum Likelihood flat--spectrum analysis developed 
by Zaldarriaga (1998).
%%%%%%%%%%%%%%%%%%%%%%
%%%%%%%%%%%%%%%%%%%%%%%%%%%%%%%%%%%%%%%%%%

In addition to measuring two low--multipole bands of the $C^E_l$
power spectrum, combining SPOrt polarization data with WMAP temperature
data will provide a measurement of the  $C^{TE}_l$ power spectrum
where errors on $T$ will be completely decorrelated from those
on $Q$,$U$.

\subsection{Measuring the reionization history of the Universe 
\label{likelihood}}

In this subsection we briefly describe our procedure to
work out cosmological parameters from SPOrt data, stressing,
in particular, our expected capacity to inspect the actual value of the
cosmic opacity $\tau$ to CMBR photons. $Q$ and $U$ maps are assumed to
be free from foreground contamination, the sensitivity worsening 
coming from this separation, estimated with the procedure suggested 
by Dodelson (1997), being already included in the quoted pixel 
sensitivity (see Table~\ref{featCMB} in Sect.~\ref{sport}). 

In particular, we shall recall and extend previous
results by Colombo \& Bonometto (2003), 
showing how measurements on 
$\tau$ and $n_s$ (primeval spectral index) are far less degenerate 
when polarization data are exploited, at variance from what happens 
if anisotropy data only are used.

SPOrt large beamwidth allows us to bin data in a fairly low number of 
pixels ($\sim 2400$ for a HEALPix parameter $Nside=16$, yielding
a mean angular distance $\sim 3.5^{\circ}$ between pixel centres); therefore, 
working in the angular (pixel) space is not numerically expensive; 
moreover this choice eases the combination of data from experiments 
with different sky coverages and detector specifications, as we
expect to have for anisotropy and polarization: anisotropy data
specify $T$ on $N_T$ pixels and polarization data specify $Q$ and $U$ 
on $N_P$ pixels. We order data into a $(N_T+2N_P)$--component {\it vector} 
${\bf x} \equiv \{T(i=1,...,N_T), Q(i=1,....,N_P), U(i=1,....,N_P)\}$.

We build artificial data sets, binned according to HEALPix, by using 
CMBFAST. Data are realizations of 
a given cosmological model 
$\mathcal M$ 
and, in each data--map, a white 
noise map in included. Therefore, the correlation matrix $\langle 
{\bf x^T}_i{\bf x}_j\rangle \equiv {\bf C}_{ij}={\bf S}_{ij}+
{\bf N}_{ij}$ (brackets mean ensemble average) is the sum of a signal 
term $\bf S$$_{ij}$, which depends on cosmology and can be evaluated 
using the relations given in Sec.~\ref{defaps}, and a term ${\bf N}_{ij}$, 
which accounts for the detector noise. We assume no correlation among 
noise in different modes and pixels, so that $\bf N$$_{ij}$ is diagonal and 
has distinct values $\sigma_T^2$ and $\sigma_P^2$, in $N_T$ and  $2N_P$ 
pixels, respectively.

Making use of the matrix ${\bf C}_{ij}$,
the likelihood of a model 
$\mathcal M^{\prime}$ 
given the synthetic data {\bf x} reads:
\begin{equation}
{\mathcal L}({\mathcal M'}|{\bf x}) = [(2\pi)^{N_T+2N_P} \det{\bf C'}]^{-1/2}
\exp[-{\bf x}^{\rm T}{\bf C'}^{-1}{\bf x}/2] ~.
\label{eq:like}
\end{equation}
Confidence regions in parameter space are found by a
likelihood ratio criterion. This procedure is applied to 
extract cosmological parameters, by inspecting sets of 
models 
${\mathcal M}'$ 
against real data.

Confidence regions 
in the $n_s$--$\tau$ plane are obtained while keeping the other parameters at
the fixed values: $\Omega_m=0.35$, $\Omega_\Lambda=0.65$, $\Omega_b=0.05$,
$h=0.65$, and assuming no tensor modes. 
Disentangling $\tau$ and $n_s$
measurements is a typical degeneracy removal allowed by polarization data.

All models considered here reionize at the nearest possible redshift,
where baryonic materials undergo a (nearly) sharp transition from
$x_e \simeq 0$ to $x_e \simeq 1$. In Sec.~\ref{sec:tau} we outlined 
that this simple reionization history, not only is not unique, but
could also be rather unlikely, if the highest values of $\tau$ found
by the first--year WMAP release are confirmed. In that section, however,
we also showed that detecting details in the reionization history, at
our expected level of sensitivity, is hard (although some indication 
is possible), while the polarization signal is grossly affected by 
the overall $\tau$ value. Hence, considering just the case of sharp 
reionization does not limit the significance of our conclusions.
We show results obtained if the model 
$\mathcal M$
has spectral index $n_s = 1$; but significant quantitative shifts 
occur only for true $n_s$ values unexpectedly different from unity.

The SPOrt sensitivity quoted in Table\ref{featCMB} corresponds to 
a pixel noise  
\begin{equation}
~~~~~~~~~~~~~~~~\sigma_P = 3.3\,  \sqrt{{2\,{\rm yr} \over \Delta t}
{1 \over E_0}}~~ \mu{\rm K}
\label{eq:pixnoise}
\end{equation}
for an $Nside=16$ pixilation;
 here $\Delta t$ and $E_0$ are
the flight duration and the detection efficiency, respectively.
Previous analysis carried out by Colombo \& Bonometto (2003) explored the
cases $\sigma_P = 4\, \mu$K, and  $\sigma_P = 2\, \mu$K. 
In this paper we report results for 
$\sigma_P = 3\, \mu$K as well, this value being closer to the SPOrt 
expected sensitivity.
For anisotropy data, assumed to originate from an 
independent experiment, we take $\sigma_T = 2\, \mu$K. A similar noise 
is typical of WMAP anisotropy data, once rescaled to our apertures. 
Cosmic and noise variances are taken into account by performing 1000
independent realizations of the model 
$\mathcal M$.
%%%%%%%%%%%%%%%%%%%%%%%%%%%%%%%%%%%%%%%%%%%%%%%%%%%%%%%%%%%%%%%%%%%
\begin{figure}
\begin{center}
\includegraphics[width=0.45\hsize,clip]{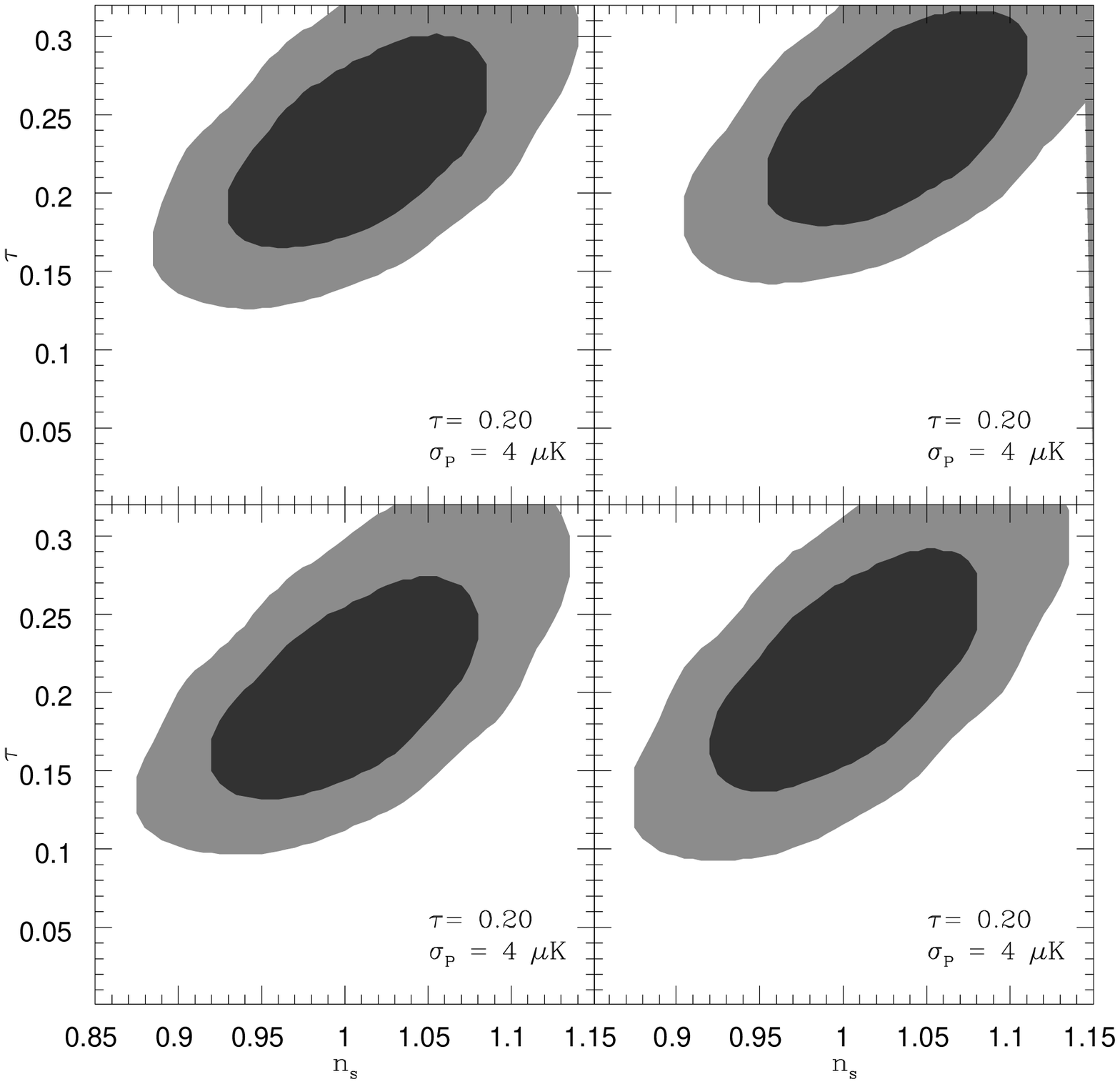}
\includegraphics[width=0.45\hsize,clip]{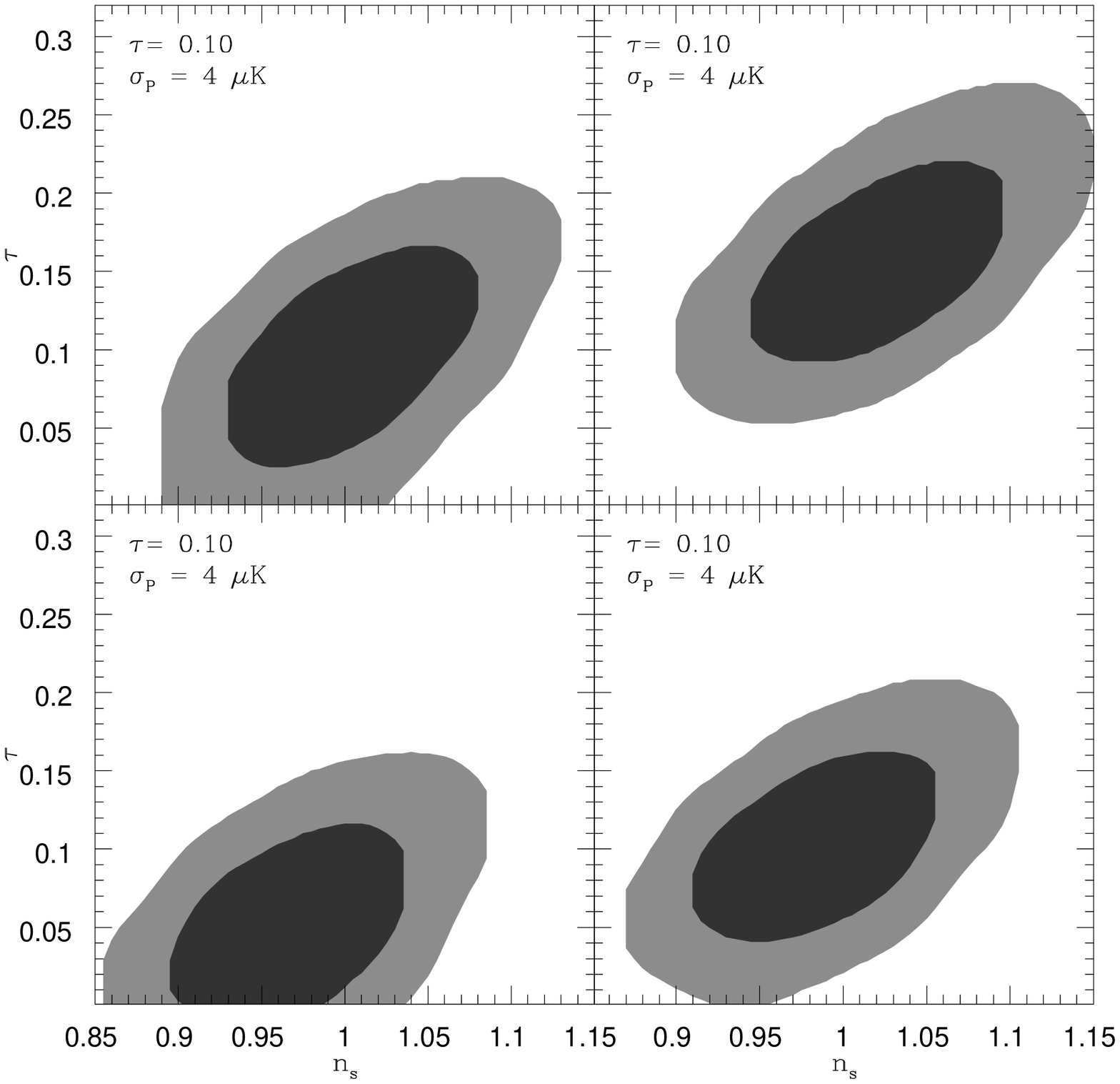}
\end{center}
\caption{Likelihood contours from the analysis of both 
polarization ($Q$ and $U$)
and anisotropy data, with $\sigma_P = 4\, \mu$K 
and $\sigma_T = 2\, \mu$K, for $\tau=0.20$ (left) and
$\tau=0.10$ (right). We assumed a flat model with
DE due to cosmological constant, $\Omega_m = 0.3$, $h=0.65$,
$\Omega_b h^2 = 0.22, n_s=1$.
Four independent realizations of the model are shown (see text).
More or less heavily shaded areas indicate 1 or 2--$\sigma$ confidence
regions, on the $n_s$--$\tau$ plane.}
\label{clt20s4}
\end{figure}
%%%%%%%%%%%%%%%%%%%%%%%%%%%%%%%%%%%%%%%%%%%%%%%%%%%%%%%%%%%%%%%%%%%%%%

\begin{figure}
\begin{center}
\end{center}
\includegraphics[width=0.45\hsize,clip]{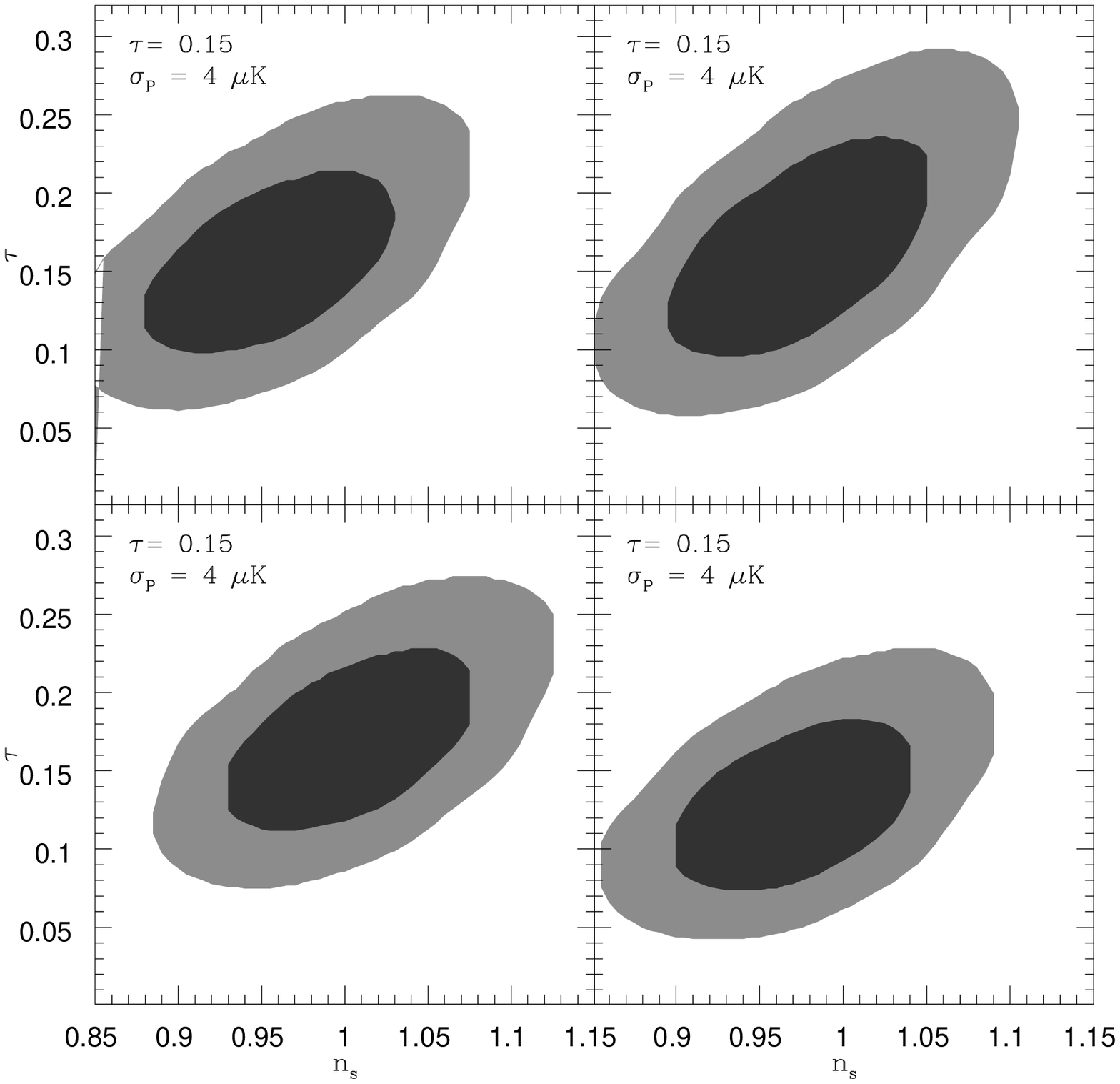}
\includegraphics[width=0.45\hsize,clip]{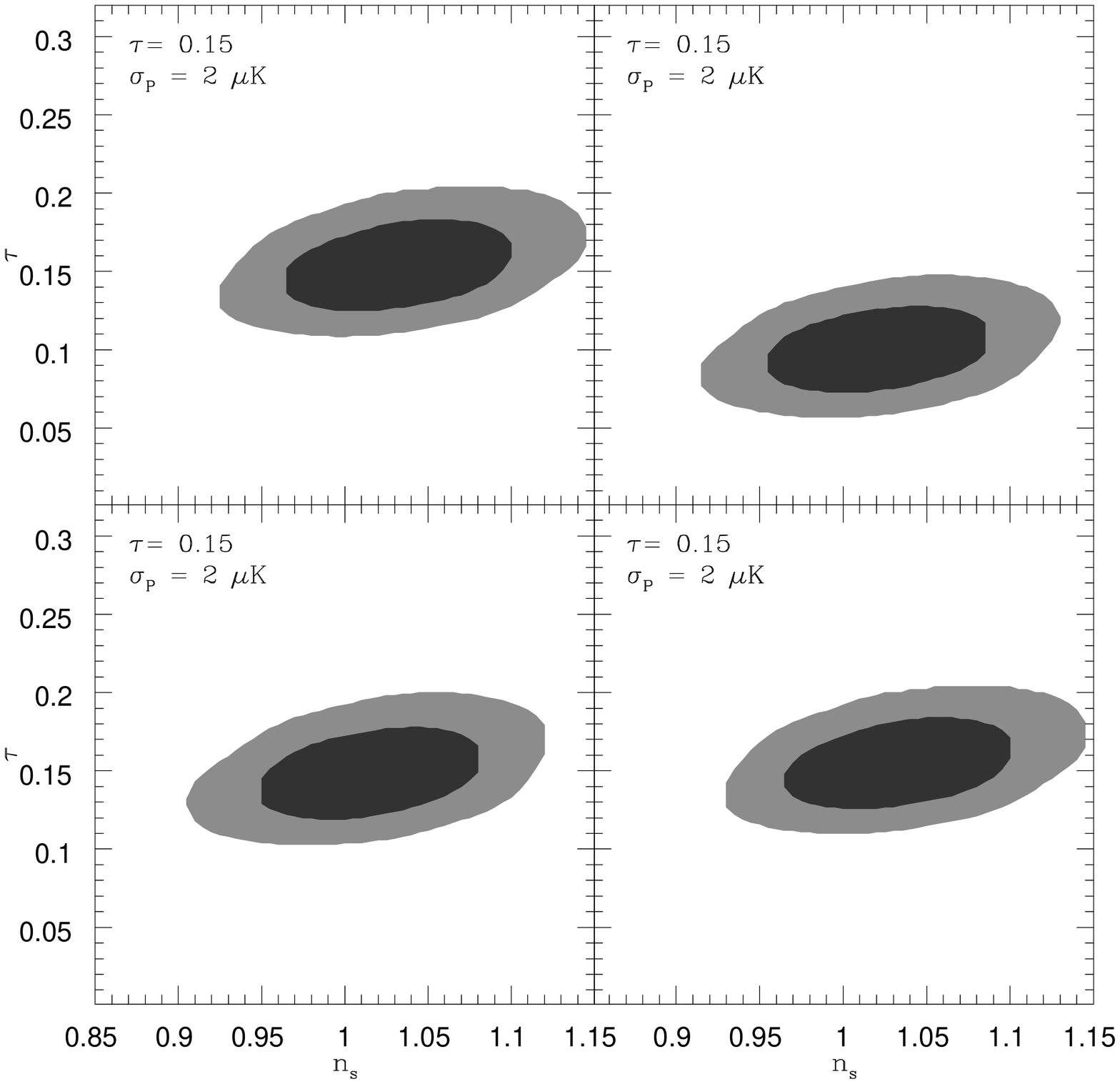}
\caption{As in Fig. \ref{clt20s4}, with $\tau=15$ and either
$\sigma_P = 4\, \mu$K (left) or $\sigma_P = 2\, \mu$K (right).}
\label{clt15ss}
\end{figure}

In Figs.~\ref{clt20s4}--\ref{clt15ss} we show joint 
confidence regions obtained in four realizations, chosen at random, for
three different values of $\tau$ in the model 
$\mathcal M$,  
with 
$\sigma_P=4\, \mu$K. as well as a realization with $\sigma_P=2\, \mu$K.

A plot similar to Figs.~\ref{clt20s4}--\ref{clt15ss} is shown by
Spergel et al. (2003), reporting WMAP results. A direct comparison
shows that we expect to constrain $\tau$ better than the
1--year WMAP release, already with $\sigma_P = 4\, \mu$K.
%%%%%%%%%%%%%%%%%%%%%%%%%%%%%%%%%%%%%%%%%%%%%%%%%%%%%%%%%%%%%%%%%%%%%%%%%
%%%%%%%%%%%%%%%%%%%%%%%%%%%%%%%%%%%%%

Let us also stress that, on very large angular scales, information on
the optical depth is carried almost exclusively by the polarization
(and cross--correlation) APS. This can be seen by comparing the
likelihood contours in Fig.~\ref{clt15ss}, where the noise on $T$ data 
is the same.

\begin{figure}
\begin{center}
\includegraphics[width=0.5\vsize,clip]{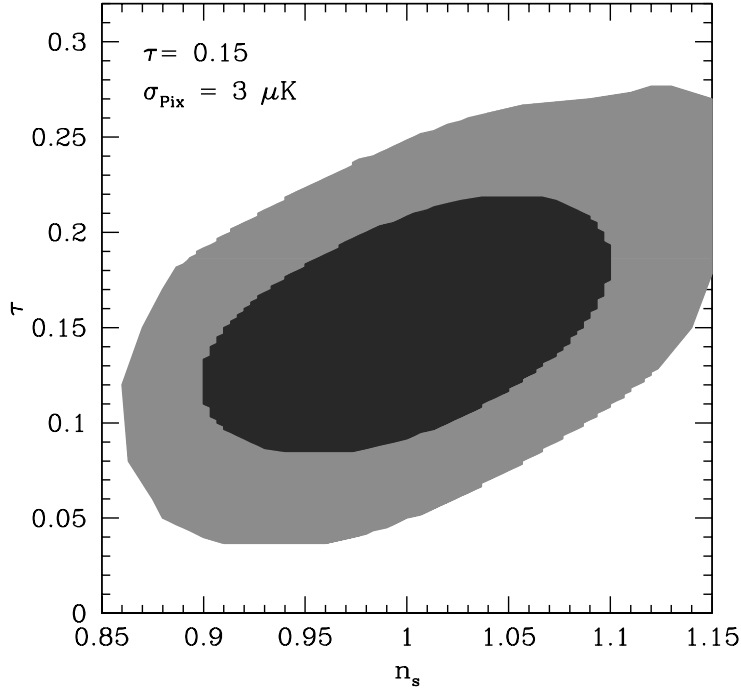}
\end{center}
\caption{Average likelihood distribution, from the analysis of 
both polarization ($Q$ and $U$) and anisotropy data, with 
$\sigma_P = 3\, \mu$K and $\sigma_T 
= 2\, \mu$K. The underlying model is the same as in Fig~\ref{clt20s4}.
 1 or 2--$\sigma$ confidence regions, in the 
$n_s$--$\tau$ plane, are shaded as in previous figures.
}
\label{avli3}
\end{figure}

In Fig~\ref{avli3} we show the likelihood contours,
averaged on cosmic and noise variances,  for $\sigma_P=3\, \mu$K.
Average results are often used to describe the
expected performance of an experiment. However, when the
likelihood distribution is not expected to approach
a Gaussian behaviour, a more detailed analysis of how 
cosmic and noise variance can
affect data is in order.
%%%%%%%%%%%%%%%%%%%%%%%%%%%%%%%%%
%%%%%%%%%%%%%%%%%%%%%%%%%%%%%%%%%%
%%%%%%%%%%%%%%%%%%%%%%%%%%%%%%%%%
\begin{figure}
\centering
\includegraphics[width=0.7\vsize,clip]{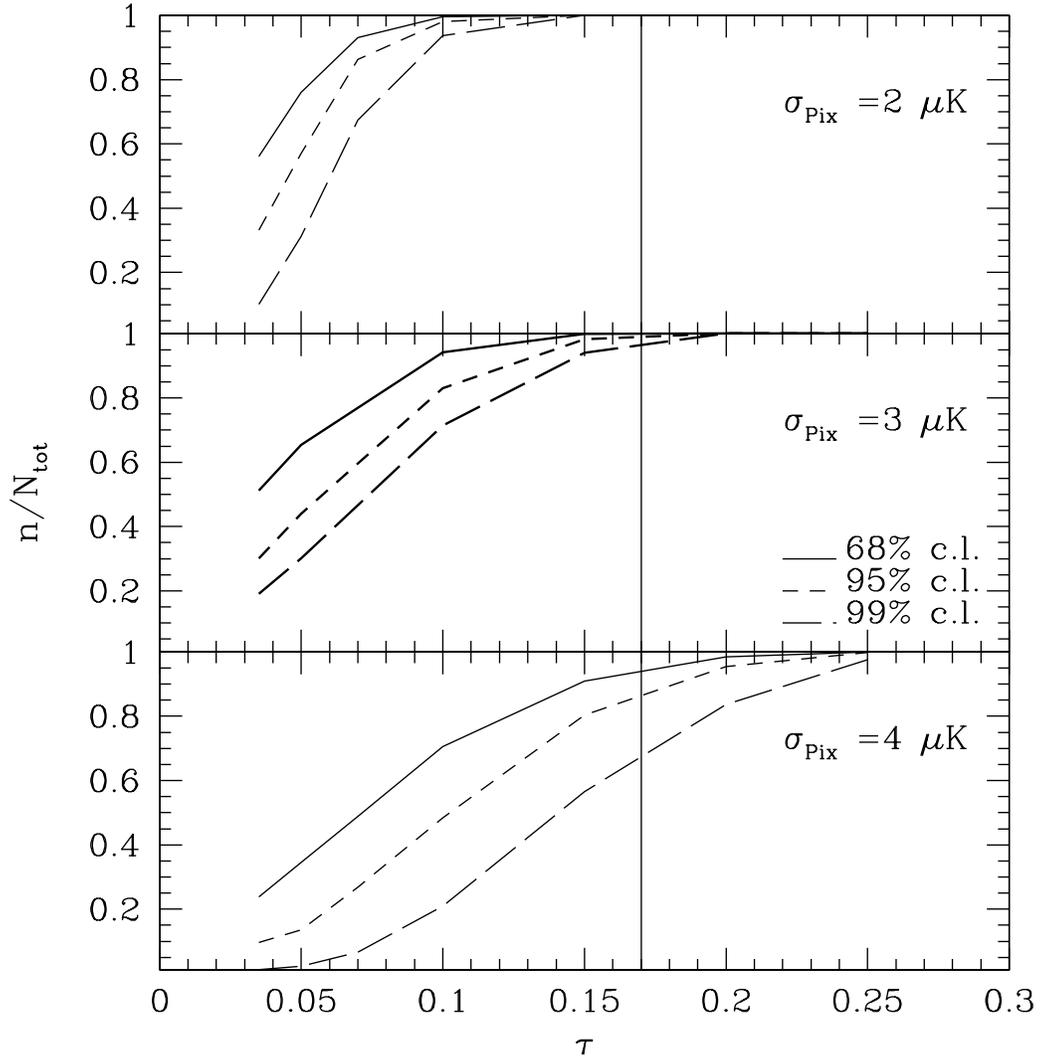}
\vskip -1.2truecm
\caption{Fractions of realizations where a lower limit on $\tau$
is found, at different confidence levels, assuming $\sigma_T = 2\, \mu$K
and three different values for $\sigma_P$.} 
\label{frac}
\end{figure}
%%%%%%%%%%%%%%%%%%%%%%%%%%%%%%%%%%
Accordingly,
Fig.~\ref{frac} shows in which fraction of realizations 
various $\tau$ limitations can be achieved. Already for $\sigma_P = 4\, \mu$K, 
minimum and maximum values of $\tau$ are detectable
in 90$\, \%$ of realizations, at 2--$\, \sigma$ level,
 provided that the real 
$\tau \ga 0.18$.
If $\tau \simeq 0.14$, such percentage reduces to $\sim 60\, \%$.
These figures should not be confused with those obtainable from the
average likelihood. In the latter case, still for $\sigma_P = 4\, \mu$K 
and at the 2--$\sigma$ level, the minimum $\tau$ value should be 
distinguishable
from zero if the real $\tau$ is just above 0.10. In this very case,
there is also a 20$\ \%$ of {\it a--priori} probability that a
minimum value of $\tau$ is fixed at the 3--$\sigma$ level.

For the case $\sigma_P = 3\, \mu$K, 
minimum and maximum $\tau$ values can be detected,
in 90$\, \%$ of realizations, at 2(3)--$\, \sigma$ C.L.,
provided that the real $\tau >\sim 0.12 (0.14)$.
Such percentage reduces to $\sim 60\, \%$ only if
the real $\tau$ is below 0.07 (0.085). Then, if we refer to the average
likelihood,  $\tau$ should be lower than 0.06
not to be distinguished from zero.

It is however worth pointing out that, even if $\tau$ was as low
as 0.03, a value smaller than the lower WMAP limit, we should still have  
a $\sim 30\, \% (20\,\%)$ probability to distinguish it from zero at  
2(3)--$\sigma$ C.L. Most of these estimates can be deduced from
Fig.~\ref{frac}, in which we also point out the case $\tau=0.17$, as
from WMAP peak probability.

Let us remind that an improvement in $\tau$ determination is certainly 
welcome, as widely discussed in Sec.~\ref{thry}. Large $\tau$ values 
were the main finding of the first--year WMAP release and the physical
implications of such $\tau$'s concern the evolution of the Universe
at the eve of object formation. Such results arise from 
the TE correlation at small $l$. Testing such correlation on data 
coming from instruments measuring $T$ and Stokes parameters separately 
is, in our opinion, extremely important and urgent.

%%%%%%%%%%%%%%%%%%%%%%%%%%%%%%%%%%%%%%%%%%%%%%%%%%%%%%%%%%%%%%%%%%%%%%%%%%%%
                                        
\section{Conclusions}
\label{conc}

The recent results of the NASA--WMAP satellite put particular emphasis on the
importance  of  CMBR--polarization investigations  on large scales, able to
either confirm or revise the estimate  $\tau=0.17$ obtained from the 
$C_{TEl}$ spectrum of the WMAP first--year data. 
Such an unexpectedly high value of $\tau$ suggests 
fascinating scenarios, some of them discussed in Sect. 
\ref{sec:tau}, like a double reionization. 
However, due to systematics still to be fully understood, WMAP did not produce any 
polarization map yet. Thus, after their first results, there are more 
and more reasons for investigating large--scale properties of CMBR 
polarization, independently from the anisotropy.

Previous attempts to measure the CMBR polarization 
 had already demonstrated that the istantaneous sensitivity of the instruments,
which essentially depends on the front--end noise, does not always reflect in
the final sensitivity. This was due to the presence of systematics
cancelling out the advantages of long integration times. 
Such systematics have been widely treated for anisotropy experiments, 
but not so deeply for polarization experiments
that led to positive detections much more recently.

The SPOrt experiment represents the first attempt to measure the $Q$ 
and $U$
Stokes parameters of the microwave sky on large angular scales
as direct outputs of the instrument. No off--line process is needed to
extract the primary information, e.g. the linear polarization in each 
pixel as defined by the $7^{\circ}$ FWHM of each SPOrt antenna.
The SPOrt radiometers have been designed to minimize every source of
systematics (in polarization) following an original analytical approach that
could lead the way to next generation instruments aiming at B--mode 
investigations of CMBR. A quite new philosophy, in fact, has been 
adopted beginning from the identification of the most critical parts
of the instrument, passing through their requirement definition and 
ending up with custom realization of components that actually represent 
a new state--of--the--art.

Following this idea several new devices have been realized, as for
example the onboard calibrator, based on a complex of marker injectors, 
that allows a full monitoring of the instrument response.
The analog correlation is 
performed by a custom passive microwave device, namely the Hybrid 
Phase Discriminator and, similarly,
other critical components of the SPOrt radiometers have been realized
reaching performances often two--three orders of magnitude better 
than equivalent off--the--shelf components.

As far as data--reduction and analysis are concerned, a new 
destriping technique has been studied to minimize the 
correlated noise due to residual systematics and
typically affecting the low order multipoles, where most of the 
cosmological information resides.
As a consequence, the SPOrt capability to measure large scale CMBR polarization
shall be similar to that of WMAP. Although it is not as high as that 
planned by PLANCK, SPOrt will make use of  {\em true} polarimeters.

In parallel to the efforts on both the analysis and the technological
development, a relevant activity has also started for understanding and
modelling the expected major source of polarized foreground, e.g. 
the Galactic synchrotron radiation.
Existing data sets on Galactic polarized emission, unfortunately limited to
regions around the galactic plane and frequencies below 2.7~GHz, have
been studied in their spectral and angular properties. Since
it was recognized that any simple extrapolation from few GHz to much higher
frequencies would fail, the strategy was to build a template 
estimating the synchrotron
polarized emission free from Faraday rotation. The result is a 
Faraday--free template whose computed APS can be extrapolated
up to 90~GHz, at least for multipoles $l<20$.

In summary, SPOrt can be expected to provide a direct measurement of 
$\tau$, discriminating it from zero at the 2--$\sigma$ level, if the 
real $\tau >\sim 0.06$. This assumes an excellent performance of
observational apparati, leading to a pixel sensitivity $\sigma_P=3\, \mu$K
for a HEALPix--like pixilation with $Nside=16$.
Actual results, of course, are subject
to cosmic and noise variances. If we consider that in more detail,
minimum and maximum $\tau$ values can be detected, in 60$\, \%$ (90$\, \%$) 
of realizations, at 2$\, \sigma$ level, if the real $\tau >\sim 0.075$
(0.12). Finally, if $\tau \simeq 0.17$, according to WMAP peak 
probability, not only considering the average likelihood, but also
in 83$\, \%$ of realizations, the detected value will be distinguishable 
from zero at the 3--$\sigma$ level.

Furthermore, at the 1--$\sigma$
level, some separate information on the reionization redshift $z_{ri}$ and
ionization rate $x_e$ may be obtainable (see Fig.~\ref{new2}). In the 
same way, we can hope 
to put some constraints on the nature of Dark Energy, thanks
to the effects on CMBR of different ISW patterns.
The SPOrt capability to obtain separate information on cosmic
reionization histories and DE nature, is however to be explored in
more detail. It should never be forgotten that the 
SPOrt outputs will be completely free of any possible bias due to 
instrumental polarization--anisotropy correlation. 

Besides testing reionization and discriminating between different values 
of the Universe optical depth, which is its most appealing goal, SPOrt 
will provide almost full sky polarization maps of the Galaxy at 22 and 
32 GHz. Last but not least, SPOrt will allow testing of
new technologies presently beyond the state of the art, thus opening a real
window upon more challenging scientific and civil space applications.

\section*{Acknowledgments}
SPOrt is a project funded by the Italian Space Agency and built by a team of
Italian industries led by Alenia Spazio.
The Italian Space Agency also gives full support to the scientific
activities carried out by the SPOrt science team.
The SPOrt team wish to thank the European Space Agency for providing
transportation and accommodation to the International Space Station as well as
for any other support given to the project since its selection in 1997.
We acknowledge use of the cmbfast and HEALPix packages.

\end{document}